\documentclass[12pt]{article}

\usepackage{fullpage}
\usepackage[T1]{fontenc}
\usepackage{ae,aecompl}
\usepackage[dvips]{graphicx}
\usepackage{amssymb}
\usepackage{amsmath}
\usepackage[round,authoryear]{natbib}
\usepackage{color}
\usepackage{array}
\usepackage{rotating}
\usepackage{colortbl}
\usepackage{tikz}
\usetikzlibrary{patterns}

\definecolor{webgreen}{rgb}{0,0.4,0}
\definecolor{webbrown}{rgb}{0.6,0,0}
\definecolor{purple}{rgb}{0.5,0,0.25}
\definecolor{darkblue}{rgb}{0,0,0.7}
\definecolor{darkred}{rgb}{0.7,0,0}
\definecolor{darkgreen}{rgb}{0,0.7,0}
\usepackage{comment}

\usepackage{hyperref}
\hypersetup{colorlinks,citecolor=darkred,filecolor=black,linkcolor=darkblue,urlcolor=webgreen}
\newcommand{\ignore}[1]{}
\newtheorem{lemma}{{\sc Lemma}}

\newtheorem{prop}{{\sc Proposition}}

\newtheorem{theorem}{{\sc Theorem}}
\newtheorem{defn}{{\sc Definition}}
\newtheorem{obs}{{\sc Observation}}

\newenvironment{proof}{\noindent {\bf \sl Proof\/}:\enspace}
{\hfill $\blacksquare{}$ \vspace{12pt}}
\usepackage{sectsty}
\sectionfont{\fontsize{14}{14}\centering\normalfont\scshape}
\subsectionfont{\fontsize{14}{14}\centering\normalfont}
\subsubsectionfont{\fontsize{12}{12}\centering\normalfont}

\usepackage{bigints}
\usepackage{enumitem}
\usepackage{cleveref}
\usepackage{mathtools}
\usepackage{caption}
\usepackage{subcaption}
\usepackage{etoolbox} 

\begin{document}
\begin{titlepage}
\title{\textbf{Optimal Robust Mechanism in Bilateral Trading}}
\author{Komal Malik\thanks{Indian Statistical institute, Delhi, India ({\tt komal.malik18@gmail.com}).} \thanks{Hebrew University of Jerusalem, Israel({\tt komal.malik@mail.huji.ac.il}).}$\;\;$ }
\maketitle

\begin{abstract}
We consider a model of bilateral trade with private values. The value of the buyer and the cost of the seller are jointly distributed. The true joint distribution is unknown to the designer, however, the marginal distributions of the value and the cost are known to the designer. The designer wants to find a trading mechanism that is robustly Bayesian incentive compatible, robustly individually rational, budget-balanced and maximizes the expected gains from trade over all such mechanisms. We refer to such a mechanism as an optimal robust mechanism.
We establish equivalence between Bayesian incentive compatible mechanisms (BIC) and dominant strategy mechanisms (DSIC). 	
We characterise the worst distribution for a given mechanism and use this characterisation to find an optimal robust mechanism. We show that there is an optimal robust mechanism that is deterministic (posted-price), dominant strategy incentive compatible, and ex-post individually rational. We also derive an explicit expression of the posted-price of such an optimal robust mechanism.
We also show the equivalence between the efficiency gains from the optimal robust mechanism (max-min problem) and guaranteed efficiency gains if the designer could choose the mechanism after observing the true joint distribution (min-max problem).

%\medskip


\end{abstract}
\thispagestyle{empty}
\end{titlepage}

\section{Introduction}
We consider a model of bilateral trading with private values. The valuation of the buyer and the cost of the seller are jointly distributed. The true joint distribution of valuation and cost is common knowledge among agents but is unknown to the designer. However, the designer knows the marginal distribution of valuation of the buyer and cost of the seller. The designer's objective is to design a mechanism that is robust to this uncertainty of the designer. In particular, we want to design a mechanism that maximizes {\sl expected welfare guarantee}. The expected welfare guarantee of a mechanism is the {\sl worst or minimum} expected gains from trade, where the minimum is taken over all joint distributions consistent with the known marginals. Also, the mechanism must be implementable for all the possible joint distributions consistent with the marginal distribution of types as agents know the true joint distribution.

Consider a setting where the designer is a third party that is designing a platform for trade and its revenue is a function of efficiency gains. It is plausible that the third party does not have precise information about the joint distribution of value and cost.  Another setting is that of a benevolent planner. In such situations, it is natural to seek distributional robustness. 
%The assumption about unknown joint distribution but knowledge about the marginal distribution of value and cost seems reasonable.
A similar approach was adopted by \cite{Carroll2017} in the monopoly setting with the buyer having multi-dimensional demand. \cite{he2022correlation} takes a similar approach in the auction environment and shows the second price auctions with no reserve price are asymptotically optimal.
\footnote{The approach adopted in the paper is one of the possible approaches and an extreme version. The set of possible joint densities could be restricted if more information is available. Also, the information to the designer can be modelled differently. For example, only moments of joint distribution could be known to designer \cite{delage2010distributionally, suzdaltsev2020essays}.}

We establish equivalence between Bayesian incentive compatible mechanisms (BIC) and dominant strategy mechanisms (DSIC). The equivalence result holds for robust efficiency gains for BIC and DSIC mechanisms along with the additional constraints on budget balancedness and individual rationality. The result implies it does not make a difference to the designer whether the joint distribution of valuations is known or unknown to the agents. The implementation of mechanisms and additional constraints over all the ``consistent" joint distributions is quite demanding and results in dominant strategy implementation with additional ex-post constraints. It is an important result as it simplifies the problem of the designer significantly. \cite{Hagerty1987} shows that a block mechanism implementable in dominant strategy, budget balanced and individually rational mechanism are implemented by ``posted-price" mechanism. This result allows us to focus on this small class of mechanisms to find an optimal robust mechanism.

The main result on the equivalence between Bayesian strategy implementation and dominant strategy implementation is observed in many other settings. \cite{bergemann2005} finds environments in which the ex-post implementation is equivalent to interim implementation for all types; the equivalence holds for separable environments, e.g., implementation of social choice function, a quasi-linear environment with no restriction on transfers. \cite{gershkov2013} and \cite{chen2019} show such equivalence in the social choice environment. \cite{manelli2010} shows the equivalence in a single unit, private values auction environment. The analysis in our paper is different from the above literature in two aspects: Firstly, we consider robustness in Bayesian incentive compatibility. Secondly, we also consider additional constraints on budget balancedness and individual rationality.    

We characterize the worst joint distribution in terms of giving the least expected welfare guarantee for a given dominant strategy incentive compatible, ex-post individually rational, and budget-balanced deterministic mechanism. We use this to show that a {\sl deterministic} posted price mechanism is an optimal robust mechanism.

Our result contributes to the growing literature on robust mechanism design pioneered by Bergemann and Morris. This literature tries to answer the Wilson critique \citep{wilson1987} of mechanisms that rely on the informational assumptions of the designer. 
\cite{bergemann2005} finds environments in which the ex-post implementation is equivalent to interim implementation for all types; the equivalence holds for separable environments, for example, implementation of social choice function, and a quasi-linear environment with no restriction on transfers. 

%In information setting similar to ours, where only marginal distributions of bidders are known, \cite{Li2020} finds that in an auction environment, second price auction with no reserve price is asymptotically optimal. The optimality of mechanism is the minimum revenue guarantee for the possible joint distributions, similar to our approach of minimum efficiency guarantee. Similar approach was adopted by \cite{Carroll2017} where the principal wants to screen an agent with multi-dimensional type with correlation. The principal just knows the marginal distribution of each component of agent's type and the mechanism is evaluated by the worst case expected profit over all the possible joint distributions.

Our paper adds to the long literature on the bilateral trading problem, which is inspired by the impossibility result in \cite{MyerEff}. It assumes independent private values, and derives the expected welfare-maximizing mechanism -- see also \cite{Chatt} for a description of equilibria of a particular bilateral trading mechanism. While these papers assume common knowledge of priors, our paper follows the models in robust mechanism design literature and relaxes this assumption.

Another strand of literature looks at robustness with respect to information structure (\citealp{Berge2017, brooks2021, carroll2018}). In those environments, there is a ``fixed" prior distribution over valuations that results from distribution over state spaces and the associated joint distribution over valuations. Then, there is an information structure that determines how the signals would be generated. The information structure affects the strategy of players as it affects the posterior beliefs about valuations. Our approach is different in the sense that only the information about priors is common knowledge, not the joint distribution. Secondly, they consider general information structures whereas we consider a particular information structure where the signal of each player reveals the true valuation of that player.

As part of the proof, we prove min-max theorem over deterministic posted price mechanisms that extends to then the class of Bayesian implementable mechanisms. It is similar to \cite{brooks2021} that proves a strong min-max theorem but for informationally robust mechanism.

The paper is organized as follows. In section~\ref{model}, we introduce the model and class of mechanisms that we are interested in. In section~\ref{result}, we present the equivalence results. Section~\ref{proof} contains the proofs of equivalence theorems, and Section~\ref{optimal} characterises the worst distribution for any (deterministic) posted price mechanism. Subsection~\ref{optResult} contains the result on optimal robust mechanism and min-max theorem. 
\section{Model}\label{model}

	We consider a private values model of bilateral trading.
	There is a single object for trade, which the seller can produce and the buyer is willing
	to buy. The valuation of the buyer for the object and the cost of the seller for producing the object
	are jointly distributed according to a distribution $H$. The marginal distribution of valuation
	of the buyer is denoted by $F,$ and the marginal distribution of cost of the seller is denoted by $G$.
	Though the true joint distribution $H$ is common knowledge among agents (the buyer and the seller), the designer
	does not know $H$.\footnote{The assumption of common knowledge about $H$ among agents is least restrictive as it allows to look at "robust" BIC mechanisms. The main result shows that we can focus on DSIC mechanisms without loss of generality. This assumption holds well in situations where the there large and number of agents who have more information about the characteristics of the good.} However, she knows the marginal distributions $F$ and $G$. 
	%We assume that the valuations and costs lie in $\theta=[0,1]$ and the marginal distributions are continuous.'
	We assume for simplification that the valuations of buyers lie in $V= [0, \overline{v}] $ and costs lie in $ C=[0, \overline{c}]$. 
	We define $\Theta= V \times C.$\footnote{In the proof of equivalence result, we prove the theorem for $\Theta= [0,1]\times[0,1]$ but it extends easily to type space with different supports (even unbounded) for marginal distributions of value of the buyer and cost of the seller.} 	
	
	We assume that marginal distribution of valuations, $F$ and $G$ are continuous i.e. there are no atoms.\footnote{Allowing for atoms does not affect the the equivalence results. %The equivalence results and the deterministic fixed price mechanism being an optimal robust mechanism continue to hold.
To see this, note that we can approximate the problem with atoms to problem with continuous marginal by spreading mass at a point to a small $\epsilon$ square; the efficiency gains from arbitrary given mechanism will be close for the two problems. }
		There are infinitely many possible joint distributions consistent with the given marginal distributions of agents. % The joint distribution does not affect the set from which mechanism can be chosen since all the properties are prior free, but 
		Note that the joint distribution affects the efficiency gains from a mechanism. 

	% Actually it extends to $[0, \infty).$
	%\debasis{There is a paper of Mezzeti and someone else
	%in JET which shows that having different support allows the impossibility in Myerson+Satt disappear
	%in some cases. This should be commented in a footnote. They call this the ``gap'' case.}: 
	
	%\cite{makowski1993possibility} 

	We focus attention on the direct revelation mechanisms.\footnote{ The designer can potentially attain information about $H$ from agents but we do not consider it as we are interested in situations where there are large number of buyers and sellers with different joint distributions.}
	A {\sl (direct) mechanism} is a triplet $(q,t_b,t_s)$, where $q:\Theta \rightarrow [0,1]$
	and $t_i:\Theta \rightarrow \mathbb{R}$ for each $i \in \{b,s\}$. Here, $q(v,c)$ denotes
	the probability of trade, $t_b(v,c)$ is the payment made {\sl by} the buyer and $t_s(v,c)$ is the
	payment made {\sl to} the seller at type profile $(v,c)$.

	\subsection{Notions of incentive compatibility}

	We introduce two notions of incentive compatibility in this section. The first two notions are standard
	in the literature.

	\begin{defn}
A mechanism $(q,t_b,t_s)$ is {\bf dominant strategy incentive compatible (DSIC)} if for every $(v,c) \in \Theta,$
\begin{align*}
v q(v,c) - t_b(v,c) &\ge v q(v',c) - t_b(v',c)~\qquad~\forall~v' \in V, \\
t_s(v,c) - c q(v,c) &\ge t_s(v,c') - c q(v,c')~\qquad~\forall~c' \in C.
\end{align*}
	\end{defn}

While DSIC is a prior-free notion, the weaker requirement of {\sl Bayesian incentive compatibility} is not.
In our model, the designer only knows the marginal distributions of types of individual agents.
Hence, we require a robust version of Bayesian incentive compatibility.

A joint probability distribution $\widehat{H}$ of $(v,c)$ is {\sl consistent} with $(F,G)$ if
the marginal distribution of $v$ and $c$ are $F$ and $G$ respectively:
\begin{align*}
\widehat{H}(v, \overline{c}) = F(v)~\qquad~\forall v \in V,\\
\widehat{H}(\overline{v}, c)= G(c)~\qquad~\forall c \in C.
\end{align*}
Let $\mathcal{H}$ denote the set of all joint distributions consistent with $(F,G)$. 
Note that the true distribution $H \in \mathcal{H}.$ Since $F$ and $G$ are continuous functions\footnote{$F$ and $G$ are absolutely continuous as continuous increasing functions are absolutely continuous.}, 
we can find well defined joint probability density function denoted by $h,$ that generates joint distribution consistent with $(F,G).$

% \begin{equation*}
%f(v)= \left\{
%\begin{array}{ll}
%\dfrac{d{F(v)}}{d v}  &\textrm{ if $F$ is differentiable at $v$}\\
%1& \textrm{ otherwise.}
%\end{array}
%\right. 
%g(c)=
%\left\{
%\begin{array}{ll}
%\dfrac{d{G(c)}}{d c} & \textrm{ if $G$ is differentiable at $c$}\\
%1& \textrm{ otherwise.}
%\end{array}
%\right.
%\end{equation*}
%By Lebesgue's Theorem for the differentiability of monotone functions, marginal distributions $F$ and $G$ are differentiable % almost everywhere. This makes $h$ consistent with $(F,G).$

\begin{defn}
A mechanism $(q,t_b,t_s)$ is {\bf Bayesian incentive compatible (BIC)} with respect to a prior $\widehat{H} \in \mathcal{H}$
if
\begin{align*}
\mathbb{E}_{c,\widehat{H}} \big[v q(v,c) - t_b(v,c)\big] &\ge \mathbb{E}_{c,\widehat{H}} \big[v q(v',c) - t_b(v',c)\big]
\qquad~\forall~v,v' \in V,\\
\mathbb{E}_{v,\widehat{H}} \big[t_s(v,c) - c q(v,c)\big] &\ge \mathbb{E}_{v,\widehat{H}} \big[t_s(v,c') - c q(v,c')\big]
\qquad~\forall~c,c' \in C,
\end{align*}
where $\mathbb{E}_{c,\widehat{H}}$ denotes the conditional expectation of $c$ given valuation $v$ using joint distribution $\widehat{H}$
and $\mathbb{E}_{v,\widehat{H}}$ denotes the conditional expectation of $v$ given cost $c$ using joint distribution $\widehat{H}.$
A mechanism $(q,t_b,t_s)$ is {\bf marginal-consistent Bayesian incentive compatible (M-BIC)} if it is BIC with respect to
{\sl all} priors $\widehat{H} \in \mathcal{H}.$
\end{defn}
Clearly, a DSIC mechanism is BIC with respect to all priors. Hence, it is M-BIC.

\subsection{Other desiderata}

It is natural to impose two additional constraints on mechanisms in the bilateral trading problem:
(a) participation constraint, and (b) budget-balance constraint.

\begin{defn}
A mechanism $(q,t_b,t_s)$ is {\bf ex-post individually rational (EIR)} if for every $(v,c)\in \Theta,$
\begin{align*}
vq(v,c) - t_b(v,c) &\ge 0, \\
t_s(v,c) - c q(v,c) &\ge 0.
\end{align*}
\end{defn}
A weaker version of individual rationality is the following.
\begin{defn}
A mechanism $(q,t_b,t_s)$ is {\bf interim individually rational (IIR)} with respect to a prior $\widehat{H} \in \mathcal{H}$
if
\begin{align*}
\mathbb{E}_{c,\widehat{H}} \big[vq(v,c) - t_b(v,c) \big] &\ge 0 \qquad~\forall~v \in V, \\
\mathbb{E}_{v,\widehat{H}} \big[t_s(v,c) - c q(v,c)\big] &\ge 0 \qquad~\forall~c \in C.
\end{align*}
A mechanism $(q,t_b,t_s)$ is {\bf marginal-consistent interim individually rational (M-IIR)} if it is IIR
with respect to all priors $\widehat{H} \in \mathcal{H}$.
\end{defn}
The M-IIR participation constraint is the analogue of M-BIC incentive constraint we had introduced earlier.
Since the designer is uncertain about the true prior, she wants to design mechanisms which satisfy these
stronger notions of IC and IR constraints. Note that these are still weaker than DSIC and EIR constraints.

We also introduce two notions of budget-balance constraints.
\begin{defn}
A mechanism $(q,t_b,t_s)$ is
\begin{itemize}
\item {\bf budget-balanced (BB)} if for all $(v,c)$, $t_b(v,c) = t_s(v,c),$
%\item {\bf Marginal-consistent budget-balanced (M-BB)} if budget balanced with respect to all priors $\widehat{H} \in \mathcal{H}$.
\item {\bf $\epsilon$-budget-balanced ($\epsilon$-BB)} given $\epsilon > 0$, if for all $(v,c)$, $|t_b(v,c) - t_s(v,c) |\le \epsilon $.
\end{itemize}
\end{defn}

The proofs construct a DSIC and EIR mechanism, but a particular type of DSIC and EIR mechanism. We call it the {\sl block mechanism}.
For this, we divide the type space $[0,1]^2$ into $n^2$ squares for any positive integer $n$.\footnote{We prove for $\Theta= [0,1]^2$ for ease of notation.} We do so in the
usual way: for each $k \in \{0,1,\ldots,n\}$, let $v_k=c_k=\frac{k}{n}$.
Then, a {\sl block} is defined as:
\begin{align*}
B_{k,\ell} &:= [v_{k-1},v_k) \times (c_{\ell-1},c_{\ell}] \qquad~\forall~k,\ell \in \{1,\ldots,n\},
\end{align*}
with the usual convention that $v_0=c_0=0$. Clearly, $B_{k,\ell}$ is different for different values of $n$.
But we suppress the dependence of $B_{k,\ell}$ on the value of $n$ for notational simplicity, unless it is necessary to be explicit.
\begin{defn}
A mechanism $(q^n,t_b^n,t_s^n)$ is an {\bf $n$-block mechanism} if for each block
$B_{k\ell}$ and for every $(v,c),(v',c') \in B_{k\ell}$, we have
\begin{align*}
q^n(v,c) &= q^n(v',c'), \\
t^n_i(v,c) &= t^n_i(v',c') \qquad~\forall~i \in \{b,s\}.
\end{align*}
\end{defn}

We will impose these notions of budget-balancedness, individual rationality, and incentive compatibility to define three classes of mechanisms.

\subsection{Three classes of mechanisms}
We consider three classes of mechanisms. These mechanisms use different notions of IC, IR, and BB constraints.
\begin{align*}
\mathcal{M}_{B} &= \{(q,t_b,t_s): ~(q,t_b,t_s)~\textrm{is M-BIC, M-IIR, and BB}\}\\
\mathcal{M}_D &= \{(q,t_b,t_s): ~(q,t_b,t_s)~\textrm{is DSIC, EIR, BB and $n$-block mechanism}\}\\
\intertext{Then, for a given $\epsilon > 0$, define}
\mathcal{M}_{\epsilon} &= \{(q,t_b,t_s): ~(q,t_b,t_s)~\textrm{is DSIC, EIR, and $\epsilon$-BB}\}.
\end{align*} 
We establish equivalence results for these classes of mechanisms in terms of robust efficiency gains\footnote{We have equivalence result between $\mathcal{M}_{\epsilon}$ and the other classes but we can easily replace $\mathcal{M}_{\epsilon}$ with mechanism that have budget surplus within $\epsilon$ margin.}.
Also, we define $\mathcal{M}_{\widehat{D}} = \{(q,t_b,t_s): ~(q,t_b,t_s)~\textrm{is DSIC, EIR, and BB}\}$ which forms an intermediate class of mechanisms and used to show equivalence between the three class of mechanisms defined above. 

As a designer, we are interested in evaluating the {\sl worst efficiency} of a mechanism in these classes.
Formally, given a mechanism $(q,t_b,t_s)$, its {\bf robust efficiency gain} is
\begin{align*}
\textsc{Eff}(q,t_b,t_s) &= \inf \limits_{\widehat{H} \in \mathcal{H}} \mathbb{E}_{(v,c), \widehat{H}} \big[(v-c)q(v,c)\big].
\end{align*}
In each classes of these mechanism, we can then define the {\sl optimal robust mechanism}. Given an $\epsilon > 0$, for any
$\mathcal{M} \in \{\mathcal{M}_{\epsilon},\mathcal{M}_B,\mathcal{M}_D\}$,
\begin{align*}
\textsc{Eff}(\mathcal{M}) &= \sup_{(q,t_b,t_s) \in \mathcal{M}} \textsc{Eff}(q,t_b,t_s).
\end{align*}

Any mechanism in $\mathcal{M}$ which attains worst case efficiency gain equal to $\textsc{Eff}(\mathcal{M})$ will be
an {\sl optimal robust mechanism} in $\mathcal{M}$.%\debasis{Does such a mechanism exist?}

\section{Equivalence results}\label{result}
There are two main equivalence results of the paper. The first theorem says that robust efficiency gain in the class
of mechanisms $\mathcal{M}_{B}$ can be made arbitrarily close to robust efficiency gain in class of
prior-free mechanisms. The precise statement is the following.

%\begin{theorem}\label{theo:main1}
%For every $\epsilon>0$, there exists $\eta(\epsilon) > 0$ and $\lim_{\epsilon \downarrow 0} \eta(\epsilon)=0$ such that
%\begin{align*}
%\textsc{Eff}(\mathcal{M}_{B}) - \textsc{Eff}(\mathcal{M}_{\eta(\epsilon)}) &\le \epsilon
%\end{align*}
%\end{theorem}

\begin{theorem}\label{theo:main1}
For every $\epsilon>0$, there exists $\eta(\epsilon) > 0$ such that
\begin{align*}
\textsc{Eff}(\mathcal{M}_{B}) - \textsc{Eff}(\mathcal{M}_{\eta(\epsilon)}) &\le \epsilon.
\end{align*}
Also, $\eta(\epsilon)$ is such that $\lim_{\epsilon \downarrow 0} \eta(\epsilon)=0.$ 
\end{theorem}

%Note that $\mathcal{M}_{\eta(\epsilon)} \subseteq \mathcal{M}_{MB}$ implies
%$\textsc{Eff}(\mathcal{M}_{MB}) - \textsc{Eff}(\mathcal{M}_{\eta(\epsilon)}) \ge 0$.

The second main theorem compares robust efficiency
gains in $\mathcal{M}_{B}$ with the class of mechanisms in $\mathcal{M}_D$.
\begin{theorem}\label{theo:main2}
$\textsc{Eff}(\mathcal{M}_{B}) = \textsc{Eff}(\mathcal{M}_{D})$.
\end{theorem}

The proof of these theorems reveal that the equivalence results are stronger than what the statements suggest. For any given mechanism in $\mathcal{M}_B$, we can find a mechanism in $\mathcal{M}_D$ that generates the gains from trade very close to that of mechanism in $\mathcal{M}_B,$ for all consistent joint distributions. As a result, it holds for \textit{worst} distribution as well. 
%Indeed, the proofs construct particular DSIC mechanisms, which we call {\sl block mechanisms}.

Since $\mathcal{M}_B \supset \mathcal{M}_{\widehat{D}}\supset \mathcal{M}_D$, Theorem~\ref{theo:main2} implies that $\textsc{Eff}(\mathcal{M}_{B}) = \textsc{Eff}(\mathcal{M}_{\widehat{D}})$. 
This establishes equivalence between the robust BIC and DSIC mechanisms with additional constraints in budget balancedness and individual rationality. 
Theorem~\ref{theo:main2} implies that we can focus on robust mechanism in the class of block mechanism implementable in dominant strategy, with ex-post individual rationality and budget balancedness. This simplifies the mechanism designer's problem significantly since \cite{Hagerty1987} showed that mechanisms in $\mathcal{M}_D$ are implementable by random posted price mechanisms, which is a much simpler class of mechanisms.

\section{Proofs of equivalence results}\label{proof}

We construct $n$- block mechanisms satisfying DSIC and EIR mechanisms to prove the equivalence theorems. 

\subsection{Proof of Theorem \ref{theo:main1}}

For proof of Theorem \ref{theo:main1}, we construct a sequence of $n$-block mechanisms starting from a
mechanism in $\mathcal{M}_{B}$.
We show that each of these block mechanisms is DSIC and EIR. The constructed block mechanism need not be BB. But, for sufficiently high value of $n$, it is arbitrarily close to budget balancedness. \footnote{In fact, the mechanism will have budget surplus for all $n$ by construction. We require close to budget balancedness for Theorem \ref{theo:main2}. }
For sufficiently high value of $n$, can come arbitrarily close to the robust efficiency gain of the original mechanism.

Given a mechanism $(q,t_b,t_s) \in \mathcal{M}_{B}$ and positive integer $n > 1$, define a new mechanism
$(q^n,t_b^n,t_s^n)$ as follows. First, we define $q^n$: for each block $B_{k,\ell}$ and for each $(v,c) \in B_{k,\ell},$
 \begin{align}
 \label{eq:all}q^n(v,c) =
\begin{cases}
 &\textrm{if}~k=1~\textrm{or}~\ell=n~\textrm{or}\\
0&~\int \limits_{B_{k-1,\ell}}q(x,y)~dx~dy = 0~\textrm{or}\\
&~\int \limits_{B_{k,\ell+1}}q(x,y)~dx~dy=0, \\
n^2 \int \limits_{B_{k,\ell}}q(x,y)~dx~dy & \textrm{otherwise}.
\end{cases}
\end{align}
The allocation probabilities in new mechanism are average probabilities in a block, except for at most $2n$ blocks.
 \begin{figure}[h!]
\centering
  \includegraphics[scale=0.8]{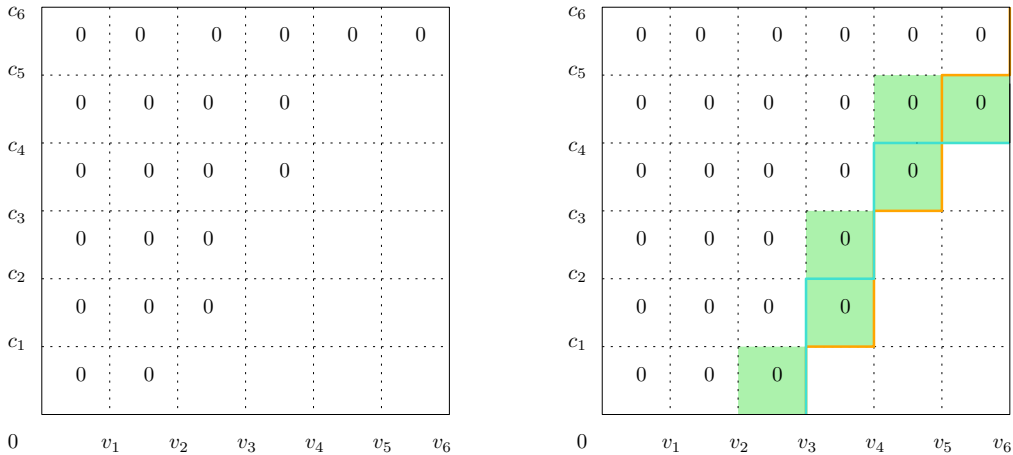}
  \caption{Relating allocations: From $q$ to $q^6$}
  \label{fig:q_n}
  \end{figure}
  
  The figure \ref{fig:q_n} illustrates how the allocation probabilities under the two mechanisms are different.
 The entry zero in block implies that the average allocation probability over the block is zero. The figure on the left has blocks with the same average probability as $q.$ From this block mechanism, we can derive block mechanism, $q^6.$ The average probabilities match for mechanism $q$ and $q^6$ on all but few blocks that coloured in green.
 The green coloured blocks have positive allocation probability for mechanism $q$ but zero allocation probability under mechanism $q^6$.

The allocation function $q^6$ weakly decreases the allocation probability for each block and increases the number of blocks with zero probability of allocation, but this happens for only at max $2n$ blocks out of $n^2$ blocks, for a general allocation function, $q^n$. As $n$ increases, the share of such blocks would be insignificant and would have the same efficiency gains as the one from just averaging out the allocation probabilities.

It is clear that $q^n$ is feasible: $n^2 \int \limits_{B_{k,\ell}}q(x,y)~dx~ dy \le n^2 \int \limits_{B_{k,\ell}}dx~ dy = 1$
for each $k,\ell$. We show in the following lemma that $q^n$ is {\sl ex-post monotone}: for each $v,v',c,c'$ with $v' > v$ and $c' < c$,
$q^n(v,c) \le q^n(v',c)$ and $q^n(v,c) \le q^n(v,c')$.

\begin{lemma}
\label{lem:ic1}
Allocation rule $q^n$ is ex-post monotone. 
\end{lemma}
\begin{proof}

\noindent {\sc Step 1.} In this step, we show that $q$ satisfies a `monotonicity' property and the
mechanism $(q,t_b,t_s)$ satisfies a version of the {\sl payoff equivalence} formula.
Fix a $v$ and $(c_{\ell-1},c_{\ell}]$. Define a joint density $\hat{h}^v$ follows:

\begin{align}
\label{eq:h}
\hat{h}^v(x,y) =
\begin{cases}
n f(v) & \textrm{if}~x=v, y \in (c_{\ell-1},c_{\ell}], \\
0 & \textrm{if}~x=v, y \notin (c_{\ell-1},c_{\ell}], \\
h(x,y) & \textrm{otherwise}.
\end{cases}
\end{align} 
where $f(v)$ is the marginal density of value of the buyer. \\

Note that $\hat{h}^v$ coincides with $h$ everywhere except for $h(v,\cdot)$.
By construction,  $ \int_{C} \hat{h}^v(v,y)dy=\int \limits_{c_{\ell-1}}^{c_{\ell}}n f(v)dy = f(v)$.
Hence, the marginals of $\hat{h}^v$ and $h$ coincide everywhere, implying $\hat{h}^v \in \mathcal{H}$.

Now consider a pair of M-BIC constraints for buyer: $v,v' \in [0,1]$ with $v > v'$.
Consider the M-BIC prior constraints of mechanism $(q,t_b,t_s)$ for prior $\hat{h}^v \in \mathcal{H}$. We have
\begin{align}
\int \limits_y v q(v,y) \hat{h}^v(v,y) dy -  \int \limits_y t_b(v,y) \hat{h}^v(v,y) dy &\ge
\int \limits_y v q(v',y) \hat{h}^v(v,y) dy -  \int \limits_y t_b(v',y) \hat{h}^v(v,y) dy \nonumber \\
\Rightarrow \int \limits_{c_{\ell-1}}^{c_{\ell}} v q(v,y) dy
- \int \limits_{c_{\ell-1}}^{c_{\ell}} t_b(v,y)  dy &\ge
\int \limits_{c_{\ell-1}}^{c_{\ell}} v q(v',y) dy
- \int \limits_{c_{\ell-1}}^{c_{\ell}} t_b(v',y) dy. \label{eq:bic1}
\end{align}
Analogously, the M-BIC constraint from $v'$ to $v$ with prior $\hat{h}^{v'}$ gives
\begin{align}\label{eq:bic2}
	\int \limits_{c_{\ell-1}}^{c_{\ell}} v' q(v',y) dy
	- \int \limits_{c_{\ell-1}}^{c_{\ell}} t_b(v',y) dy &\ge
	\int \limits_{c_{\ell-1}}^{c_{\ell}} v' q(v,y) dy
	- \int \limits_{c_{\ell-1}}^{c_{\ell}} t_b(v,y) dy.
\end{align}
Adding (\ref{eq:bic1}) and (\ref{eq:bic2}) gives
\begin{align*}
(v - v') \int \limits_{c_{\ell-1}}^{c_{\ell}} \big[q(v,y) - q(v',y) \big]dy &\ge 0.
\end{align*}
Since $v > v'$, we get for all $\ell \in \{1,\ldots,n\}$,
\begin{align}\label{eq:mon1}
\int \limits_{c_{\ell-1}}^{c_{\ell}} q(v,y) dy &\ge \int \limits_{c_{\ell-1}}^{c_{\ell}} q(v',y) dy.
\end{align}
Analogously, for each $c < c'$, we get for all $k \in \{1,\ldots,n\}$,
\begin{align}\label{eq:mon2}
\int \limits_{v_{k-1}}^{v_{k}} q(x,c) dx &\ge \int \limits_{v_{k-1}}^{v_{k}} q(x,c') dx.
\end{align}We use this to prove ex-post monotonicity of $q^n$ in the next step. 

Next, using (\ref{eq:bic1}) gives us for every $v$ and every $\ell$,
\begin{align*}
u_b(v,\ell) \ge \int \limits_{c_{\ell-1}}^{c_{\ell}} v q(v',y) dy
- \int \limits_{c_{\ell-1}}^{c_{\ell}} t_b(v',y) dy,
\end{align*}
where 
\begin{align}
  \label{eq:IC}  u_b(v,\ell):= \int \limits_{c_{\ell-1}}^{c_{\ell}} v q(v,y) dy
- \int \limits_{c_{\ell-1}}^{c_{\ell}} t_b(v,y)  dy.
\end{align} 
Rewriting
\begin{align*}
u_b(v,\ell) &\ge u_b(v',\ell) + (v-v') \int \limits_{c_{\ell-1}}^{c_{\ell}} q(v',y) dy.
\end{align*}
Hence, $u_b(\cdot,\ell)$ is convex in the first argument for every $\ell$ and
$\int \limits_{c_{\ell-1}}^{c_{\ell}} q(v',y) dy$ is the subgradient of $u_b(\cdot,\ell)$
at $v'$. By the fundamental theorem of calculus, we can thus write for every $v$ and every $\ell$
\begin{align}\label{eq:payoff1}
u_b(v,\ell) &= u_b(0,\ell) + \int \limits_0^v \Big(\int \limits_{c_{\ell-1}}^{c_{\ell}} q(x,y) dy \Big)dx.
\end{align}
This payoff equivalence formula is useful and will be used later. \\

\noindent {\sc Step 2.} Fix $v \in [v_{k-1},v_k)$ for some $k$. If $q^n(v,c) > 0$, for $c'$ with $c' > c$, we argue that $q^n(v,c) \ge q^n(v,c')$.
Suppose $(v,c) \in B_{k,\ell}$ and
$(v,c') \in B_{k,\ell'}$. If $\ell=\ell'$, we are done since $q^n(v,c')=q^n(v,c)$. So, $c' > c$ implies
$\ell' > \ell$. Then,

\begin{align*}
\frac{1}{n^2}q^n(v,c') &\le \int \limits_{c_{\ell'-1}}^{c_{\ell'}} \int \limits_{v_{k-1}}^{v_{k}} q(x,y) dx dy \le
\int \limits_{c_{\ell-1}}^{c_{\ell}} \int \limits_{v_{k-1}}^{v_{k}} q(x,y) dx dy = \frac{1}{n^2}q^n(v,c),
\end{align*}
where we used (\ref{eq:mon2}) for the second inequality. 

Now, suppose $q^n(v,c)=0$ for some $c$. We show that $q^n(v,c')=0$ for all $c' > c$.
Suppose $(v,c) \in B_{k,\ell}$ and
$(v,c') \in B_{k,\ell'}$. If $\ell=\ell'$, we are done since $q^n(v,c')=q^n(v,c)$.
Else, $\ell' > \ell$. Hence, $\ell \ne n$. Since $q^n(v,c)=0$, this means $q^n(v,c'')=0$ for
all $c''\in (c_{\ell},c_{\ell+1}]$. If $\ell'=\ell+1$, we are done. Else, we repeatedly apply
this procedure to get $q^n(v,c')=0$.

This shows that $q^n(v,c) \ge q^n(v,c')$ for all $c < c'$ and all $v$. An analogous proof using (\ref{eq:mon1})
can be done to show that $q^n(v,c) \le q^n(v',c)$ for all $v' > v$ and for all $c$.
\end{proof}

By Lemma \ref{lem:ic1}, since $q^n$ is monotone, we can define an EIR and DSIC
mechanism using standard revenue equivalence techniques: payments at the lowest types
are set to zero and local incentive constraints bind to give payments at all types.
In particular, for every $(v,c) \in B_{k,\ell}$,

\begin{align}\label{eq:payb}
t^n_b(v,c) =
\begin{cases}
0 & \textrm{if}~k=1, \\
t^n_b(v_{k-2},c_{\ell}) + v_{k-1} \big[ q^n(v_{k-1},c_{\ell}) - q^n(v_{k-2},c_{\ell})\big]& \textrm{otherwise.}
\end{cases}
\end{align}
Similarly, for every $(v,c) \in B_{k,\ell}$,
\begin{align}\label{eq:pays}
t^n_s(v,c) =
\begin{cases}
0 & \textrm{if}~\ell=n, \\
t^n_b(v_k,c_{\ell+1}) + c_{\ell} \big[ q^n(v_k,c_{\ell}) - q^n(v_k,c_{\ell}+1)\big]&  \textrm{otherwise.}
\end{cases}
\end{align}

\begin{figure}[h!]
\centering
\begin{subfigure}[h]{0.45\textwidth}
  \includegraphics[scale=0.9]{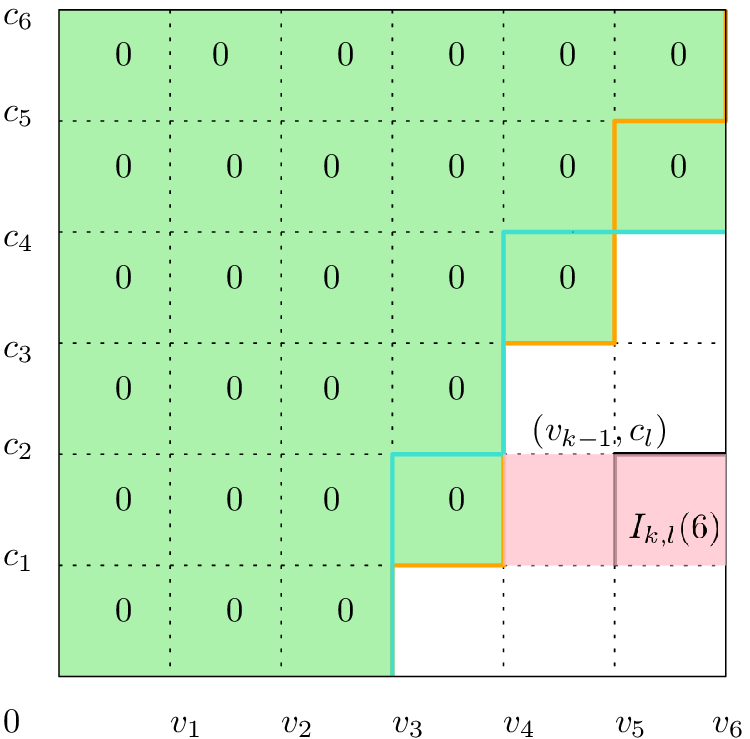}
  \caption{Payment of the buyer}
\end{subfigure}
  \hfill
\begin{subfigure}[h]{0.45\textwidth}
    \includegraphics[scale=0.9]{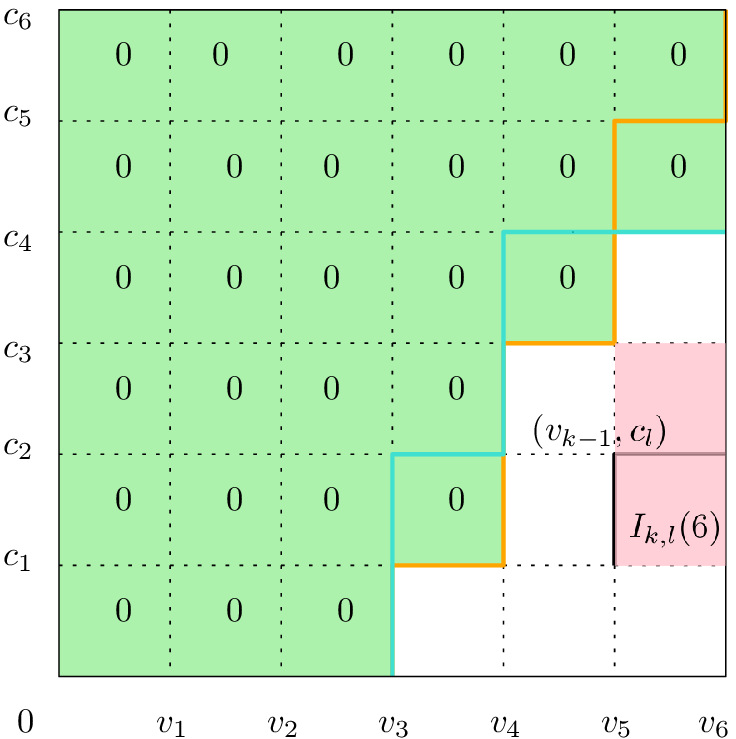}
    \caption{Payment to the seller}
    \end{subfigure}
    \caption{Relating payment rules: $(t_b, t_s)$ and $(t_b^6, t_s^6)$}
    \label{fig:pay}
  \end{figure}

The payment of buyer and seller is such that the agents with valuation at boundary of squares in the block are indifferent between reporting their true valuation and misreporting valuation in next block as shown in the figure \ref{fig:pay}. The corresponding blocks are shown in pink colour. 
Thus, we have a DSIC and EIR mechanism $(q^n,t_b^n,t_s^n)$ and gives the next lemma.

\begin{lemma}\label{lem:ic}
$(q^n,t_b^n,t_s^n)$ is DSIC and EIR mechanism.
\end{lemma}

%\begin{lemma}
%\label{lem:bb}
%For every $\eta>0$, there exists $n^*$ such that for all $ n>n^*$, the mechanism, $(q^n, t^n_b , t^n_s)$ in the sequence is such that 
%$|t_b^n(v,c) - t_s^n(v,c)|< \sup_{(v,c)}| t_b(v,c)- t_s(v,c)| + \eta,~ \forall (v,c) \in \Theta.$
%\end{lemma}

As $n$ grows infinitely, the budget surplus for a pair $(v,c)$ approaches that of original mechanism, $(q, t_b, t_s) \in \mathcal{M}_B$ which is budget balanced. We get the following lemma as a result. 
\begin{lemma}\label{lem:bb}
As $n \rightarrow \infty , $ we have $\lim_{n \rightarrow \infty } |t_b^n(v,c) - t_s^n(v,c)|=0$ for all $(v,c)\in \Theta.$
\end{lemma}

\begin{lemma}
\label{lem:got}
The efficiency gains generated by sequence of mechanism $(q^n, t^n_b, t^n_s)$ converges to efficiency gains from mechanism $(q,t_b,t_s)$, $\forall \widehat{H} \in \mathcal{H}.$
\end{lemma}

\begin{proof}
%We show that efficiency gains generated from the mechanism $(q^n, t^n_b, t^n_s)\in \mathcal{M}_{\eta}$ converges to efficiency gains from mechanism $(q,t)$, $\forall \widehat{H} \in \mathcal{H}.$
 
\noindent The efficiency gains of mechanism $(q^n, t^n_b, t^n_s)$ is given by
\begin{align*}
 \int \limits_{\Theta} (v-c)~ q^n(v,c)~d\widehat{H}(v,c)  
 =& \sum_{k, \ell} ~\int \limits_{ B_{k, \ell}(n)} (v-c)~ q^n(v,c)~d\widehat{H}(v,c).
\end{align*}
%We would show the convergence between the two expressions. \\
We would use the following lemma directly at this point. It will be proved in Section \ref{Appendix}.
\begin{lemma}
\label{lem:conv2}
For any block $B_{k, \ell}(w)$, we have
\begin{align}
\label{eq:conv_vp} \lim_{m \rightarrow \infty} \bigintssss \limits_{B_{k, \ell}(w)} (v-c)~q^{wm}(v,c) ~d\widehat{H}(v,c) = \bigintssss \limits_{B_{k, \ell}(w)}(v-c) q(v,c)~d\widehat{H}(v,c) + \dfrac{2}{w } \bigintssss \limits_{B_{k, \ell}(w)} q(v,c)~d\widehat{H}(v,c),
\end{align}
where $q^{wm}$ is allocation for block mechanism with $(wm)^2$ blocks.
\end{lemma}

Summing over $(k, \ell)$ in equation (\ref{eq:conv_vp}), we get
\begin{align*}
\nonumber \sum_{k, \ell  }\bigg|  \bigintssss \limits_{B_{k, \ell}(w)}& (x-y)~q^{wm}(x,y)~ d\widehat{H}(x,y) - \bigintssss \limits_{B_{k, \ell}(w)} (x-y)~q(x,y) ~d\widehat{H}(x,y)\bigg|\\
&\le\dfrac{2}{w} \sum_{k, \ell} \left[  \int \limits_{B_{k, \ell}(w)}  q(x,y) ~d\widehat{H}(x,y)\right]\\
 &\le \dfrac{2}{w}.
 \end{align*}
The last inequality follows from the fact that $\forall (x,y), q(x,y) \le 1$ and $\widehat{H}$ is a probability measure.
Thus, as $w \rightarrow \infty,$ and $m \rightarrow \infty,$ we have 
\begin{align*}
 \sum_{k, \ell  } \bigintssss \limits_{B_{k, \ell}(w)}& (x-y)~q^{wm}(x,y)~ d\widehat{H}(x,y) \rightarrow \sum_{k, \ell} \bigintssss \limits_{B_{k, \ell}(w)} (x-y)~q(x,y) ~d\widehat{H}(x,y).
 \end{align*}
Note that we had partitioned the entire space of valuation into $(wm)^2$ squares. Choosing $w=m= \sqrt{n}$, we can see that for all $\widehat{H} \in \mathcal{H}$, as $n \rightarrow \infty$, efficiency gains of $(q^n, t^n_b, t^n_s)$ converges to efficiency gains of mechanism $(q,t_b, t_s).$
\end{proof}

By combining Lemma \ref{lem:ic}, Lemma \ref{lem:bb} and Lemma \ref{lem:got},  for all $\widehat{H} \in \mathcal{H}$ and for all $(q, t_s, t_b) \in \mathcal{M}_{B}$, 
\begin{align*}
\mathbb{E}_{(v,c), \widehat{H}} \big[(v-c)q(v,c)\big]&=  \lim_{n \rightarrow \infty} \mathbb{E}_{(v,c), \widehat{H}} \big[(v-c)q^n(v,c)\big] \textrm{ and }
\lim_{n \rightarrow \infty} |t^n_b(v,c)- t^n_s(v,c)|=0. 
\end{align*}
For every $\epsilon>0$, for all $\widehat{H} \in \mathcal{H},~(q, t_s, t_b) \in \mathcal{M}_{B}$, there exists $\eta(\epsilon)$ such that $\lim_{ \epsilon \downarrow 0} \eta(\epsilon)=0$, $(q^n, t^n_b, t^n_s) \in \mathcal{M}_{\eta(\epsilon)}$ for large $n,$ and 
\begin{align*}
\mathbb{E}_{(v,c), \widehat{H}} \big[(v-c)q(v,c)\big]&\le \mathbb{E}_{(v,c), \widehat{H}} \big[(v-c)q^n(v,c)\big]+ \epsilon. 
\end{align*}
Taking infimum over $\widehat{H} \in \mathcal{H}$, we get that for every $\epsilon>0$ and for all $(q, t_s, t_b) \in \mathcal{M}_{B}$, there exists $\eta(\epsilon)$ such that $\lim_{ \epsilon \downarrow 0} \eta(\epsilon)=0$ and for large $n$, we have $(q^n, t^n_b, t^n_s) \in \mathcal{M}_{\eta(\epsilon)}$ and 
\begin{align*}
\textsc{Eff}(q,t_b,t_s)& \le  \textsc{Eff}(q^n,t_b^n,t_s^n) + \epsilon. 
\end{align*}
Taking supremum over all $(q,t_b,t_s) \in \mathcal{M}_{B}$ and using the fact that $(q^n , t^n_b, t^n_s) \in \mathcal{M}_{\eta(\epsilon)}$, we get that for every $\epsilon>0$, there exists $\eta(\epsilon)$ such that $\lim_{ \epsilon \downarrow 0} \eta(\epsilon)=0$ and 
\begin{align*}
\textsc{Eff}(\mathcal{M}_{B})&\le \textsc{Eff}(\mathcal{M}_{\eta(\epsilon)}) + \epsilon.
\end{align*}

\subsection{Proof of Theorem \ref{theo:main2}}
%We start by showing that for every $\epsilon > 0$, there exists $\eta(\epsilon) > 0$ and $ \lim_{\epsilon \downarrow 0} \eta(\epsilon)=0$ such that

We define class of mechanisms, $\mathcal{M}_{\eta}^B:=\{(q^n , t_b^n, t_s^n): (q,t_b, t_s) \in \mathcal{M}_{B},(q^n , t_b^n, t_s^n) \in \mathcal{M}_{\eta} \}.$ Here, $(q^n, t_b^n, t_s^n)$ is constructed using (\ref{eq:all}), (\ref{eq:payb}) and (\ref{eq:pays}) from mechanism $(q, t_b, t_s).$

\begin{lemma}\label{lem:dsic}
As $\eta \downarrow 0$, we have
$\textsc{Eff}(\mathcal{M}^B_{\eta}) \rightarrow \textsc{Eff}(\mathcal{M}_{D}).$
\end{lemma}

%Any mechanism $(q_0, t_0) \in \mathcal{M}_D$ is budget balanced whereas $(q,t) \in \mathcal{M}_B$ has budget surplus within $\eta$. As $\eta$ gets very small, the difference between allocation probabilities also gets very small and even if the allocation probability, $q$ may be giving higher gains of trade, it bounded and cannot be very different from gains of trade from $q_0.$ Since it happens for all mechanisms, in the corresponding classes, the difference between robust efficiency gains also diminishes, implying that robust efficiency gains from $\mathcal{M}_{\eta \downarrow 0} $ can be bounded above by robust gains of trade in $\mathcal{M}_D.$\\

\begin{proof} We start by mentioning two results for $n$-block mechanisms implementable in dominant strategy. The proof of the following observations is provided in Section \ref{Appendix}.
\begin{obs}\label{ob1}
For an $n$-block mechanism $(q,t_b,t_s)$ implementable in dominant strategy, for $k,l \in \{ 1,2,\dots, n-1\}$ we have
\begin{multline}
t_b(v_k, c_{\ell})-t_s(v_k, c_{\ell}) \\ =(v_k- c_{\ell}) q(v_k, c_{\ell}) -\dfrac{1}{n} \left[\sum_{i=0}^{k-1} q(v_i, c_{\ell})
+ \sum_{i=l+1}^{n} q(v_k, c_i) \right]+ t_b(v_0, c_{\ell})- t_s(v_k, c_n).
\end{multline}
\end{obs}

\begin{obs}\label{ob2}
The constructed $n$-block mechanism, $(q^n, t^n_b, t^n_s)$ has the property that for all $n$, for any $k< \ell,$ we have $q^n(v_k, c_{\ell})=0$. 
\end{obs}

\cite{Hagerty1987} shows that a block mechanism $(q, t_b, t_s)$ which is DSIC, BB and EIR is implementable by posted price mechanism. In particular, such a mechanism with $n$ blocks can be represented by a vector $\mathbf{u}_{(n-1) \times 1}$ where $i^{\textrm{th}}$ element, $u_{i}= q (v_{i}, c_i )$ where $\sum_{i=1}^{n-1} u_i \le 1.$ \footnote{See Corollary 3 in \cite{Hagerty1987} for details.}
Also, for any $k<l$, we have $q(v_l, c_k)=0.$

Consider a constructed $n$-block mechanism $M=(q^n, t_b^n,  t_s^n) \in \mathcal{M}^B_\eta$. Here $n$ is the smallest number such that for every $n'\ge n$, the constructed $n'$-block mechanism lies in $\mathcal{M}^B_\eta.$ We construct from $M$ a block mechanism $(q_0, t_0, t_0)$ which is DSIC, BB and EIR using the result in \cite{Hagerty1987}. Let vector $\mathbf{u}$ be the corresponding vector for the block mechanism, $(q_0,t_0,t_0)$ and be defined as
\begin{align*}
u_i= q^n(v_{i}, c_i) + \Delta_n, ~~~i \in \{ 1, \dots, n-1\}.
\end{align*}
Here $\Delta_n$ is a carefully chosen constant. We will define it later. 

\noindent We show that allocation function $q_0$ is well defined and $ \lim_{\eta \rightarrow 0 } \left[ q^n(v,c) - q_0(v,c) \right]=0$ in three steps.\\

\noindent {\sc Step 1.} The allocation function $q_0$ is well defined iff $\sum_{i=1}^{n-1} u_i \le 1.$
From Observation~\ref{ob1}, for $k>\ell$, we have \footnote{For the concerned block mechanisms, $t_b^n(v_0, c_{\ell})=t_0(v_0, c_{\ell}) = t_s^n(v_k, c_n)= t_0(v_k, c_n)=0.$}
 \begin{align*}
q^n(v_k, c_{\ell}) &= \dfrac{1}{(v_k- c_{\ell})}\left( t_b(v_k, c_{\ell})-t_s(v_k, c_{\ell}) +\dfrac{1}{n} \left[\sum_{i=0}^{k-1} q^n(v_i, c_{\ell})
+ \sum_{i=l+1}^{n} q^n(v_k, c_i) \right]\right).
%q_0(v_k, c_{\ell}) &= \dfrac{1}{(v_k- c_{\ell})}\left( \dfrac{1}{n} \left[\sum_{i=1}^{k-\ell} q_0(v_i, c_{\ell})
%+ \sum_{i=l+1}^{n} q_0(v_k, c_i) \right]\right)\\
 \end{align*}
 
\noindent Iteratively replacing the expression of allocation rules on the right hand side of the above equation, and using Observation~\ref{ob2}, we get
 \begin{align}
\label{eq:iterative}q^n(v_k, c_{\ell}) &= r_n(k, \ell)+ \sum_{i=\ell}^{k}q^n(v_{i}, c_i),
\end{align}
where
\begin{equation*}
%r_n(k, \ell)= \dfrac{n}{k-\ell} \bigg( t^n_b(v_k,c_{\ell}) - t^n_s(v_k, c_{\ell})\\
%\left. + \dfrac{1}{k-\ell-1} \left[ t^n_b(v_{k-1},c_{\ell}) - t^n_s(v_{k-1}, c_{\ell} )+  t^n_b(v_{k},c_{\ell+1}) - t^n_s(v_{k}, c_{\ell+1} )+ \dfrac{1}{k-\ell-2}\left(\dots \right) \right] \right).\\
r_n(k,\ell)=\left\{
\begin{array}{ll}\dfrac{n}{k-\ell} \left( t^n_b(v_k, c_\ell) - t_s^n(v_k, c_\ell) + \dfrac{1}{n}  \sum_{i=\ell+1}^{k-1} \left(r_n(k,i)+ r_n(i, \ell) \right) \right) & \textrm{if } k>\ell+1,\\
n(t^n_b(v_k, c_\ell) - t_s^n(v_k, c_\ell))& \textrm{if } k=\ell+1.
\end{array}
\right.   
\end{equation*}

 We choose $\Delta_n=  \dfrac{r_n(n,1)}{n-1}.$ Note that $\displaystyle  \sum_{i=1}^{n-1} u_i=   \sum_{i=1}^{n-1} q^n(v_i, c_i) + r_n(n, 1) = q^n(v_n,c_1).$ The first equality follows from the definition of $\mathbf{u}$ and the second equality follows from equation \eqref{eq:iterative}.
 Since $q^n(v_{n} ,c_1)\le 1, $ it must be true that $ \sum_{i=1}^{n-1} u_i \le 1$. \\

 \noindent {\sc Step 2.} 
We find the expression for $q^n(v,c) - q_0(v,c) $ for all $(v,c) \in \Theta.$ Fix $n$ and a utility profile $(v,c) \in B_{k, \ell}(n).$ Note that $k, \ell$ corresponding to $v,c$ depend on $n$. \\

\noindent {\sc Case 1. } For $(v, c) \in B_{k,l}(n)$ where $k < \ell$ for all $n:$ From Observation~\ref{ob2}, it follows that $ ~q^n(v,c)=0.$ In \cite{Hagerty1987}, the construction from vector $\mathbf{u}$ is such that $q_0(v,c)=0.$ Thus, $q^n(v,c)- q_0(v,c)=0.$\\

\noindent {\sc Case 2.} For $(v,c) \in B_{k,k}(n)$: By construction of $q_0$, we have 
\begin{align}
   \label{eq:29} q^n(v,c)- q_0(v,c)= -\Delta_n= -\dfrac{r_n(n,1)}{n-1}.
\end{align}

\noindent {\sc Case 3.}
For $(v,c)\in B_{k,l}(n)$ where $k> \ell: $ As $k > \ell$, we have $ v>c$.
We then get
 \begin{align}
\nonumber q^n(v,c)- q_0(v,c)&= r_n(k, \ell) - (k-\ell+1) \Delta_n\\
 \label{eq:qn-q0}&= r_n(k, \ell) - \dfrac{k-\ell+1}{n-1} r_n(n,1).
 \end{align}

\noindent {\sc Step 3.} We show that $\lim_{\eta \downarrow 0} [q^n(v,c) - q_0(v,c) ]= 0.$ \\

%Next, we show $\lim_{\eta \downarrow 0} [q^n(v,c) - q_0(v,c) ]= 0.$ for {\sc Case 3.}
If $n$ is bounded as $\eta \rightarrow 0$, it is easy to see that $\lim_{\eta \downarrow 0}\overline{r}_n(k, \ell)=0$ as $k$, $\ell$ and $n$ are finite.\\

Let us consider the situation where $n \rightarrow \infty$ as $\eta \rightarrow 0$. Note that we will have $k$ and $\ell$ such that $\lim_{n \rightarrow \infty}  \frac{k}{n}=v$ and $\lim_{n \rightarrow \infty}\frac{\ell}{n}= c$.

We use the following observations to finding the bound on value of $r_n(k', \ell')$ where $k'> \ell'$ are arbitrary. In particular, we show that $\lim_{\eta \downarrow 0}r_n(k', \ell')=0$. 
\begin{obs}\label{obs3}
    For $k'> \ell'$, we have
\begin{align}
 \label{eq:boundonr}r_n(k', \ell')\le \dfrac{k'-\ell'+1}{2}n~ \max_{i,j} \left[t_b^n(v_i, c
_j)- t_s^n(v_i, c_j)\right] \eqqcolon \overline{r}_n(k', \ell').
\end{align}
\end{obs}
We provide proof of Observation~\ref{obs3} in Section \ref{Appendix}.

\begin{obs}\label{obs4}
    For all $v>c,$ we have
\begin{align*}
\lim_{n \rightarrow \infty}  \left[t^n_b(v,c)- t^n_s(v, c)\right]=\lim_{n \rightarrow \infty }n^2\int \limits_{c-\frac{1}{n}}^c\int \limits_{v}^{v+ \frac{1}{n}}\left[ t_b(x,y)- t_s(x,y) \right]~ dx ~dy=0.
\end{align*}
\end{obs}
Observation \ref{obs4} is derived as equation \eqref{eq:24} in the proof of Lemma \ref{lem:bb}.\\

By Observation \ref{obs4}, for all $v>c,$ we have
\begin{align}
   \label{eq:25} \lim_{n \rightarrow \infty} n^2\left[  t^n_b(v,c)- t^n_s(v, c)\right]&=\lim_{n \rightarrow \infty }n^4\int \limits_{c-\frac{1}{n}}^c\int \limits_{v}^{v+ \frac{1}{n}}\left[ t_b(x,y)- t_s(x,y) \right]~ dx ~dy=0.
\end{align}
 By Observation \ref{obs3}, we have
\begin{align}
\nonumber \overline{r}_n(k', \ell')&\le n^2~ \max_{i,j} [t_b^n(v_i, c
_j)-t_s^n(v_i, c_j)],\\
\nonumber \lim_{n \rightarrow \infty} \overline{r}_n(k', \ell')&\le \lim_{n \rightarrow \infty} n^2~ \max_{i,j} [t_b^n(v_i, c_j)-t_s^n(v_i, c_j)]=0.
\end{align}
The last equality follows from \eqref{eq:25}. However, since $r_n$ is non-negative, we get
\begin{align}
  \label{eq:26}  r_n(k', \ell')=0.
\end{align}
We use equation \eqref{eq:26} to find an upper bound on $ q^{n}(v,c)- q_0(v, c)$. For {\sc Case 2,} from equation \eqref{eq:29}, we have

   \begin{align}
\nonumber \lim_{\eta \downarrow 0} q^{n}(v,c)- q_0(v, c)&= \lim_{n \rightarrow \infty} \dfrac{r_n(n,1)}{n-1}=0,
\end{align} 
which follows from equation \eqref{eq:26}.\\

\noindent For {\sc Case 3,} from equation \eqref{eq:qn-q0}, we have
\begin{align}
\nonumber \lim_{\eta \downarrow 0} \left[ q^{n}(v,c)- q_0(v, c)\right]&= \lim_{n \rightarrow \infty} \left[r_n(k, \ell)- \dfrac{k-\ell+1}{n-1}r_n(n,1)\right].
\end{align}
Since equation \eqref{eq:26} holds for all $k'> \ell'$, we get
\begin{align}
\nonumber \lim_{\eta \downarrow 0} \left[q^{n}(v,c)- q_0(v, c)\right]&=0 + \left( \lim_{n \rightarrow \infty} \dfrac{k-\ell+1}{n-1} \right) \cdot 0.
\end{align}
Using the fact that $\lim_{n \rightarrow \infty} \frac{k}{n}= v$ and $\lim_{n \rightarrow \infty} \frac{\ell}{n}= c,$ we get
\begin{align}
\label{eq:27}\nonumber \lim_{\eta \downarrow 0} [q^{n}(v,c)- q_0(v, c)]&= (v-c) \cdot 0=0.
\end{align}
Thus, we have shown that $\lim_{\eta \rightarrow 0} [q^n(v,c) - q_0(v,c)] = 0$ for all $(v,c) \in \Theta. $
This further implies that for all $ \widehat{H} \in  \mathcal{H}$, for all $(q^n, t_b^n, t_s^n)$, as $\eta \rightarrow 0$, we have
\begin{align*}
\mathbb{E}_{(v,c), \widehat{H}} \big[(v-c)q^n(v,c)\big]& \rightarrow  \mathbb{E}_{(v,c), \widehat{H}} \big[(v-c)q_0(v,c)\big]. 
\end{align*}
Thus, for all $(q^n, t_b^n, t_s^n)$, as $\eta \rightarrow 0$, we have
\begin{align*}
 \textsc{Eff}(q^n, t_b^n, t_s^n)&\le \textsc{Eff}(q_0,t_0,t_0).
 \end{align*}
 This implies
 \begin{align*}
\lim_{\eta \rightarrow 0 } \textsc{Eff}(\mathcal{M}^B_\eta)&\le \textsc{Eff}(\mathcal{M}_D).
\end{align*}
\end{proof}

\noindent By the proof of Theorem \ref{theo:main1} and Lemma \ref{lem:dsic}, we get
\begin{align*}
\textsc{Eff}(\mathcal{M}_{B})&\le \textsc{Eff}(\mathcal{M}_D).\\
\end{align*}
But $\textsc{Eff}(\mathcal{M}_{B})\ge \textsc{Eff}(\mathcal{M}_D)$ since $\mathcal{M}_B \supseteq \mathcal{M}_D$. This implies $\textsc{Eff}(\mathcal{M}_{B})= \textsc{Eff}(\mathcal{M}_D).$

\section{Optimal robust mechanism}\label{optimal}

%By Theorem \ref{theo:main2} and  \cite{Hagerty1987},  WLOG we can restrict our attention to dominant strategy implementable mechanisms with desirable properties. 

We introduce additional notations to ease the analysis. Instead of using marginal distributions, we would use marginal density of valuation from now on as it is easier to work with. 
\begin{comment}From the given marginal distribution functions, we can determine the density functions as follows\footnote{By Lebesgue's Theorem for the differentiability of monotone functions, marginal distributions $F$ and $G$ are differentiable almost everywhere. As a result, the efficiency gains value would be unaffected by the choice of $f$ and $g$ at the points where the marginal distributions are not differentiable.}:
\begin{equation*}
f(v)= \left\{
\begin{array}{ll}
\dfrac{d{F(v)}}{d v}  &\textrm{ if $F$ is differentiable at $v$}\\
1& \textrm{ otherwise.}
\end{array}
\right. 
\end{equation*}
\begin{equation*}
g(c)=
\left\{
\begin{array}{ll}
\dfrac{d{G(c)}}{d c} & \textrm{ if $G$ is differentiable at $c$}\\
1& \textrm{ otherwise.}
\end{array}
\right.
\end{equation*}
%\begin{defn}
%Let $\mathcal{H}^d$ be the set of all the joint densities consistent with marginal densities, $f(v)$ and $g(c).$
%Mathematically,
%$$ \mathcal{H}^d= \left\{ h(v,c): f(v)= \int_c h(v, c)~ dc  \textrm{ and }  g(c)= \int_v h(v,c)~dv, \forall (v,c)\in \Theta  \right\} $$
%\end{defn}
\end{comment}
Since marginal distributions of value and cost are continuous, there exist marginal density functions of value of buyer and cost of seller, and are denoted by $f(v)$ and $g(c)$, respectively.

A joint probability density $\hat{h}$ of $(v,c)$ is {\sl consistent} with $(f,g)$ if
the marginal density of $v$ and $c$ are $f$ and $g$ respectively:
\begin{align*}
 \int_c \hat{h}(v, c)~ dc=f (v)~\qquad~\forall v  \in V,\\
\int_v \hat{h}(v, c)~ dv= g(c)~\qquad~\forall c \in C.
\end{align*}
Let $\mathcal{H}^d$ denote the set of all joint densities consistent with $(f,g)$.\\

%\noindent {\bf \sc Objective of designer : }  Consider an arbitrary budget balanced mechanism, $(q,t) \in \mathcal{M}.$ The trade probability and the true joint probability density, $h$ will determine the efficiency of mechanism and is given by 
%$ \mathbb{E}_{(v,c), h} \big[(v-c)q(v,c)\big]$. 
We can redefine the objective of designer in terms of joint density.  The designer would find a mechanism that maximises robust efficiency gains in a class of mechanisms where
\textit{robust efficiency gains} of a mechanism $(q,t)$ is given as
\begin{align*}
\textsc{Eff}(q,t) &= \inf \limits_{\hat{h} \in \mathcal{H}^d} \mathbb{E}_{(v,c), \hat{h}} \big[(v-c)q(v,c)\big].
\end{align*}

%For any mechanism, efficiency depends on joint probability density of valuations which is unknown to designer.  A natural measure of efficiency of a mechanism would be the worst efficiency, which is refereed to as \textit{robust efficiency gains} of a mechanism $(q,t)$ and is given as
%\begin{align*}
%\textsc{Eff}(q,t) &= \inf \limits_{\hat{h} \in \mathcal{H}^d} \mathbb{E}_{(v,c), \hat{h}} \big[(v-c)q(v,c)\big]
%\end{align*}
%The designer would find a mechanism that maximises robust efficiency gains in a class of mechanisms. 

Recall that $\mathcal{M}_{\widehat{D}}$ is the class of mechanisms satisfying dominant strategy incentive compatibility, budget balancedness and ex-post individual rationality. \cite{Hagerty1987} shows that mechanism in class of block mechanisms, any mechanism in $ \mathcal{M}_{\widehat{D}}$ can be implemented by posted price mechanisms- randomisation of the deterministic posted price mechanisms. %In the previous chapter, we showed that if we are interested in maximising robust efficiency gains, it is without loss of generality to consider block mechanism satisfying DSIC, BB and EIR. 
This combined with Theorem \ref{theo:main2} implies that the designer can focus without loss of generality on the class of posted price mechanisms.

Let $\mathcal{M}_P$ be the collection of all posted price mechanisms. The objective of the designer is to find optimal mechanism $(q^*,t^*)$, where
\begin{align*}
\textsc{Eff}(\mathcal{M}_P) &:= \sup_{(q,t) \in \mathcal{M}_P} \textsc{Eff}(q,t)= \textsc{Eff}(q^*,t^*).
\end{align*}
%\footnote{Note that we require $q(\cdot)$ to be integrable w.r.t to measure $H(v,c) \in \mathcal{H}(F,G)$, for the the problem to be well defined.  }

\subsection{Worst distributions}\label{Char}
Now we restrict our attention to deterministic posted price mechanisms. In such mechanisms, there is a posted price $p$. If the valuation of buyer, $v$ is greater than the posted price, $p$ and valuation of seller, $c$ is less than posted price, $p$, then the trade occurs with certainty and price, $p$ will be charged as payment. If the valuation of buyer, $v$ is less than the posted price, $p$ or valuation of seller, $c$ is greater than posted price, $p$, then trade does not occur with certainty and no payment is made.

\noindent A deterministic posted price mechanism, $M^p= (q^p, t^p)$ is defined as follows
\begin{equation*}
q^p(v,c)= \left\{ 
\begin{array}{ll}
1&\textrm{ if }  v> p \textrm{ and } c< p\\
0&\textrm{ if }  v< p \textrm{ or } c> p\\
\end{array}
\right.;
\end{equation*}
\begin{equation*}
t^p(v, c) = \left\{ 
\begin{array}{ll}
p&\textrm{ if }  v> p \textrm{ and } c< p\\
0&\textrm{ if }  v< p \textrm{ or } c> p\\
\end{array}
\right..
\end{equation*}
When $v=p$ or $c=p$, the tie between trading and not trading can be broken in anyway we want.

The efficiency gains of $M^p$ for true joint probability density $h$ is given by 
  \begin{align}
  \nonumber  \int_{v>p} \int_{c<p} (v-c)~h(v, c)~ dc ~dv 
        &=\int_{v>p} \int_{c<p} v\cdot h(v, c)~ dc ~dv  -   \int_{v>p} \int_{c<p} c \cdot h(v, c)~ dc ~dv\\
   \label{eq:gains}   &=\int_{v>p}  v\left[\int_{c<p}h(v, c)~ dc\right] ~dv  -   \int_{c<p} c\left[ \int_{v>p}h(v, c)~ dc \right]~dv.
   \end{align}
   
%The efficiency gains is a non-negative number for posted price mechanism. Since we are not considering distributions with mass points, the functional value of $q^p(v,c )$ and $t^p(v,c)$ where one of the agent has valuation as $p$, does not affect efficiency gains. 

\begin{figure}[h!]
\centering
\includegraphics[scale=1]{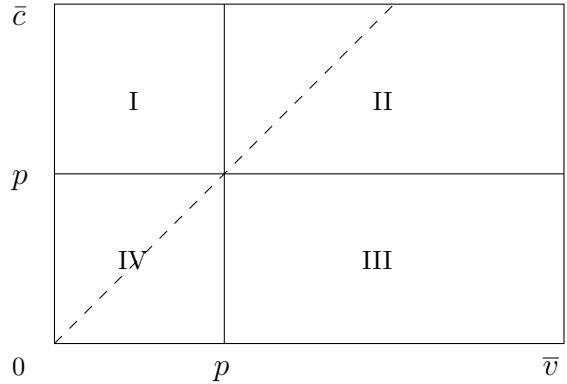}
\caption{Trade region }
\label{fig:region}
\end{figure}

Consider Figure \ref{fig:region}. The value of the buyer and the cost of the seller is represented horizontally and vertically, respectively. Given a posted price mechanism $M^p$, the trade occurs only in region III of Figure \ref{fig:region}, and will be referred as trade region. Note that the efficiency gains of a mechanism depends only on the marginal distribution of value of buyer and cost of seller within the trade region.  %Notice that the worst distribution would be such that the mass in trade region is minimum and is distributed such that the mass is closer to the $45^{\circ}$ degree line.

For a given deterministic posted price mechanism, we characterise the set of joint distributions satisfying the marginal distribution of valuations that minimises the efficiency gains. To do that, we would use the concept of redistributing the mass which is explained in the next section.  

 \subsubsection{Redistribution of mass}
Consider a joint probability density, $\hat{h} \in  \mathcal{H}^d$ such that rectangles $A$ and $D$ have same mass, say $m$ in the corresponding regions. We introduce the idea of redistribution of mass from $A$ and $D$ to $B$ and $C$. The redistribution would reduce the mass in regions $A$ and $D$ to zero whereas the mass in regions $B$ and $D$ will increase by mass, $m$.  \\
\begin{figure}[h!]
\centering
 \includegraphics{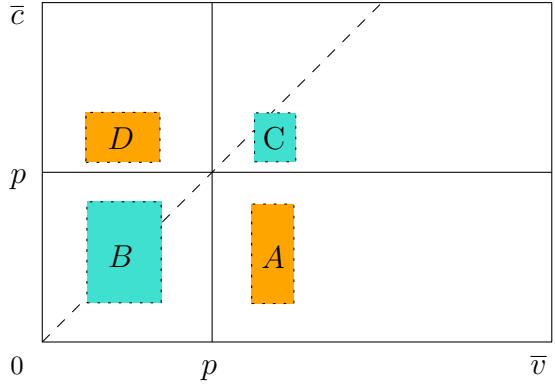}
 \caption[scale=1.2]{Redistributing mass}
 \label{fig:redist}
 \end{figure}
 
 %Define for  $M \in \{ A,B,C,D\}$, 
 %\begin{align*}
% x_M&= \min_{(x,y) \in M} \{ x\}\\
 % y_M&= \min_{(x,y) \in M} \{ y\}
 %\end{align*}
 %Consider $\Delta(x,y)= \min\{f(v_A + x, c_A+y), f(v_D + x, c_D+y)\}$
% The new joint density function, $h'(v,c)$ is defined as follows :\\

For a given $\hat{h}$, consider a new joint probability density $h'$
 \begin{equation}
 \label{eqn1}
 h'(v,c)= \left\{ 
 \begin{array}{cl}
0
& \textrm{ if } (v,c) \in A \cup D,\\\\
\dfrac{ \int_{y:(v,y) \in B \cup D} \hat{h}(v,y) ~dy \int_{x:(x,c) \in B \cup A} \hat{h}(x,c)~dx}{\int_{(x,y) \in B \cup A} \hat{h}(x,y)~dx~dy} & \textrm{ if }(v,c) \in B,\\\\
\dfrac{\int_{y:(v,y) \in  C\cup A} \hat{h}(v,y) ~dy \int_{x:(x,c) \in C \cup D}\hat{ h}(x,c)~dx}{\int_{(x,y) \in C \cup D} \hat{h}(x,y)~dx~dy} & \textrm{ if }(v,c) \in C,\\\\
\hat{h}(v,c) & \textrm{ otherwise.}
 \end{array}
 \right.
 \end{equation}

The above joint probability density is constructed such that marginal density of a valuation restricted to just the rectangles $A, ~B, ~C,$ and $D$ collectively remains unchanged. Since $\hat{h} \in \mathcal{H}^d$, we get $h' \in  \mathcal{H}^d.$ To be more specific, the marginal density of $c$ over rectangle $A$ and marginal density of $v$ over rectangle $D$ is shifted to rectangle $B$.  The marginal density of $c$ over rectangle $D$ and marginal density of $v$ over rectangle $A$ is shifted to rectangle $C$. This gives us the marginal densities over each of the rectangles. 

Given the marginal densities over a rectangle, we can construct a joint probability density function where the random variables are independent. Suppose $\alpha(x)$ and $\gamma(y)$ are the marginal densities over a rectangular region, $R \equiv X \times Y$. A possible joint probability density, where the random variables are independent in rectangular region is given by $\alpha(x)~\gamma(y)/\int_X \alpha(x) dx\equiv \alpha(x)~\gamma(y)/\int_Y \gamma(y)dy $. Using this fact, we get $h'(v,c).$ 

Notice that such a redistribution will decrease the mass in trade region and reduce the efficiency gains of mechanism $M^p$. We will use the redistributions of the above form to characterise the worst distribution for a given posted price mechanism.  
\subsection{Characteristics of worst distribution for a given posted price mechanism}\label{worst}
%Consider a given joint density function, $h(v,c)$ satisfying the given marginal densities. Corresponding to it, 

Fix mechanism $M^p.$ We define $\ell_p :=\int_p^{\overline{v}}  f(v) ~dv- \int_{p}^{\overline{c}} g(c)~dc$. It is the minimum mass from a joint distribution that must lie in the trade region in order to meet the requirement of mass imposed by the marginal distributions of valuations. Note that $\ell_p$ is a parameter and depends on just the given marginal distributions.
%We define $a \equiv \int_{0}^p g(c)~dc- \int_{0}^p f(v)~dv$.
%For $a \le 0, $ the gains from trade for the distribution can be zero.
 %Now, we consider the marginals such that $  \int_{0}^p g(c)~dc- \int_{0}^p f(v)~dv >0.$
 
\begin{figure}[h!]
\centering
\includegraphics[scale=1.2]{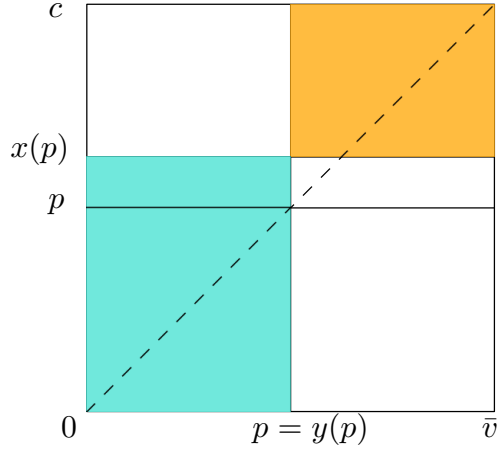}%
\caption{Worst distribution given posted price mechanism with price $p$}
\label{img1}
\end{figure}

If $\ell_p\le0,$ there is a possibility where the entire mass can be distributed such that mass in region corresponding to efficiency gains is zero. This possibility is shown in Figure \ref{img1}.
%\begin{figure}[h]
%\label{fig:lp}
%\centering
%\includegraphics[scale=0.65]{/Users/komalmalik/GoogleDrive/ISI/ISI/phd/basic_dist.png}%
%\caption{Worst distribution}
%\end{figure}
%The region for valuations is divided into four regions, based on whether valuation lies below or above $p$. The numbers in the region represent the mass in the region. \\\\
%On the given regions, such a distribution is possible. Though, we need to explicitly state the joint density that would satisfy the above distribution of mass. We define it below:
%\begin{equation}
% \label{eqn3}
% h(v,c)= \left\{ 
% \begin{array}{cl}
%\dfrac{f(v)~g(c)}{\int_{p}^{\bar{c}} g(c)~dc}& \textrm{ if } (v,c) \in B\\\\
%\dfrac{f(v)~g(c)}{\int_{0}^{p} f(v)~dv}& \textrm{ if } (v,c) \in C\\\\
%f(v) g(c)\left(1- \dfrac{\int_0^p g(c) dc}{\int_{0}^{p} f(v)~dv}\right) \left( 1-\dfrac{\int_p^{\overline{v}}f(v)~dv}{\int_{p}^{\overline{c}} g(c)~dc} \right)& \textrm{ if } (v,c) \in A\\\\
%0 & \textrm{ otherwise.}
 %\end{array}
 %\right.
% \end{equation}

Since efficiency gains is a non-negative number, for the given posted price mechanism, the robust efficiency gain must be 0.

Now we characterise the worst distribution when $\ell_p>0$. Note that if this is the case, a distribution depicted in Figure \ref{img1} is not feasible. Thus, any distribution consistent with given marginal distribution of valuations must have positive mass in trade region. 

\begin{figure}[h!]
\centering
\includegraphics[scale=1.2]{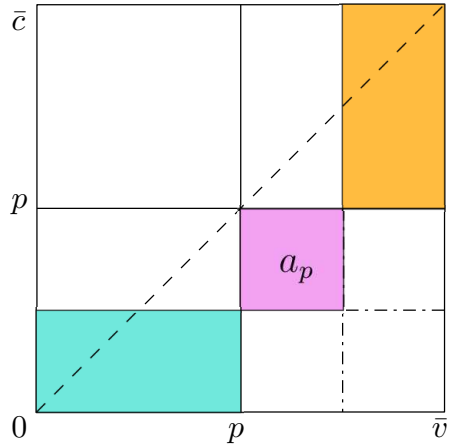}
\caption{Mass in trade region}
\label{fig:lem1}
\end{figure}

Let $\hat{h}^p$ be the worst distribution for mechanism $M^p$. We define $\displaystyle a_p:= \int \limits_p^{\bar{v}}  \int \limits_0^p \hat{h}^p(x,y) ~dy ~dx. $
For $\ell_p>0, $ we must have $a_p>0$. Such a situation is depicted in Figure \ref{fig:lem1}.

We prove few useful lemmas that characterise the worst case distribution, given a deterministic posted price mechanism, $M^p$. 

\begin{lemma}\label{lem1}
If $a_p>0$, then $z:= \int \limits_0^{p}  \int \limits_p^{\bar{c}} \hat{h}^p(x,y)~ dy~ dx=0 $. 
\end{lemma}
\begin{proof}
Suppose not for contradiction. If $a_p>0$ and $ z >0$, then we can find rectangle $A$ and rectangle $D$ with mass $m>0 $  as shown in Figure \ref{fig:lem1b}. \\
\begin{figure}[h!]
\centering
 \includegraphics[scale=1.1]{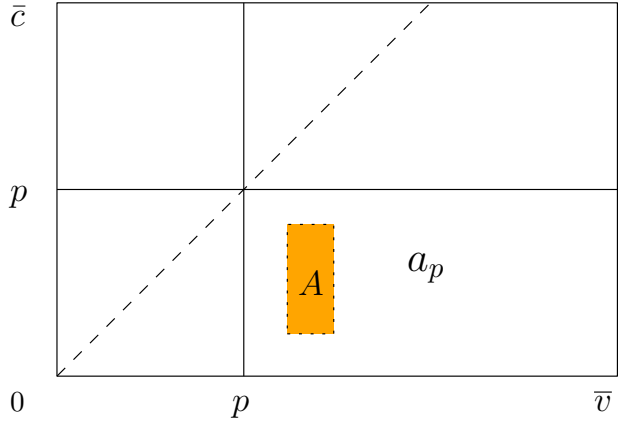}
 \caption{Revenue decreasing redistribution}
 \label{fig:lem1b}
 \end{figure}\\
Redistribute mass from $A$ and $D$ to $B$ and $C$ in the manner described by equation (\ref{eqn1}). As a result, efficiency gains will decrease, contradicting the assumption that we started with worst distribution.
\end{proof}

To further characterise the worst case distribution, we define the set $G$ as the smallest open rectangle with vertex $(p,p)$ and diagonally opposite vertex, $(v,c)$ where $v>p,$ and $c<p$, such that $\int_{(x,y)\in G}\hat{h}^p(x,y) ~dy ~dx= a_p. $\\\\
Formally, 
let $R_{k,\ell} \equiv (p, k) \times (\ell,p)$. We have
$ G \equiv \textrm{int} \left\{ \cap_{(k,\ell )\in T  } R_{k,\ell} \right\} $ where
$T= \{(k,l): \int_p^k \int_{\ell}^p \hat{h}^p(x,y)~dx~dy =a_p  \}.$\\

\noindent We define two rectangles, 
\begin{align*}
L_G&=\{(v,c): \exists (x,c) \textrm{ such that } (x,c) \in G, ~v \in [0,p)\},
\end{align*}
and
\begin{align*}
U_G&= \{(v,c): \exists (v,y) \textrm{ such that } (v,y) \in G, ~c\in (p, \overline{c}) \}.
\end{align*}

\begin{figure}[h!]
\centering
\includegraphics{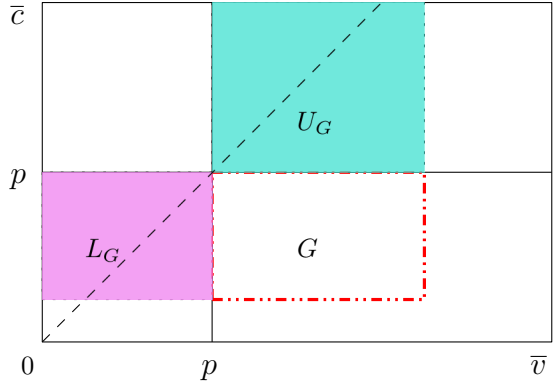}
\caption{Zero mass in left and upper region to $G$}
\label{fig:G}
\end{figure}

\noindent $L_G$ is the rectangle to the left of $G$ and $U_G$ is a rectangle upward $G$ as depicted in the Figure~\ref{fig:G}. The dotted red rectangle shows the boundary of set $G$. In the next lemma, we show that mass of area to the left, $L_G$ and above, $U_G$ is zero.

\begin{lemma}\label{lem2}
For worst distribution $\hat{h}^p,$ mass in region $L_G$ and $U_G$ is zero, i.e., 
 \begin{equation*}
      \int_{L_G} \hat{h}^p(x,y)~dx~dy=\int_{U_G} \hat{h}^p(x,y)~dx~dy=0.
 \end{equation*}
\end{lemma}

\begin{proof}
We start with the proof for $L_G$ region.
 
 \begin{figure}[h!]
 \centering
\includegraphics{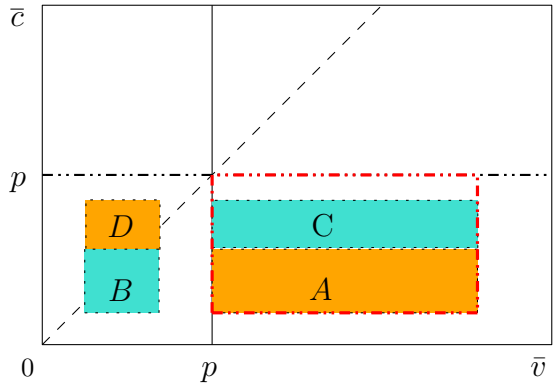} 
\caption{Revenue decreasing redistribution for $L_G$ region}
\end{figure}
 
\noindent Suppose for contradiction that there exists a rectangle, with non-zero mass $m$ in region $L_G.$ Let $D'$ be the rectangle in $L_G$ with mass, $m$.\footnote{Without loss of generality, we could have considered $D'= L_G$.}
% $$  D'=  \textrm{int} \left\{ \cap_{  T \in \mathcal{T}} T \right\}$$
% where $\mathcal{T}=\{  T : T \textrm{is a rectangle }, T \subseteq [0,p]\times [0, p] \textrm{ and} \int_{T} h(x,y) ~dx~dy=m\} $\\\\
 \noindent Now, consider rectangle $A'$ defined as follows:
 \begin{align*}
     A'=\{(v,c) : c \in \{ x: (x,y) \in D'\}\} \cap G.
 \end{align*}
Since $D'$ has positive mass, $A' \ne \emptyset.$
Note that we can find $D \subseteq D'$ and $A \subseteq A'$ such that $D$ and $A$ have same mass, $\{c: (v, c) \in A\} \cap \{c: (v,c \in D)\}= \emptyset$, and $\inf D' \ge \sup A'.$ To find these $A$ and $D$, one can find a horizontal line cutting regions $A'$ and $D'$. As the horizontal line moves upward, the mass below the line in rectangle $A'$ will increase and mass above the line in rectangle $D'$ will decrease.\footnote{For any line passing through $G$, the mass of region $A$ is strictly positive; it follows from definition of $G$. } As a result, one can choose horizontal line for region close to the bottom boundary of set $G$ and keep it moving upward until the mass equalises in region above the horizontal line in rectangle $D'$ and region below horizontal line in rectangle $A'$.
% should have better explanation

Now consider redistribution from  $A $ and  $D $ to $B$ and $C$ as per equation (\ref{eqn1}). Notice that as a result, the efficiency gains will decrease, contradicting the assumption that we started with worst distribution. 
Analogously, we could argue for upper rectangle $U_G$. The redistribution corresponding to this case is shown in Figure \ref{fig:ug}:

 \begin{figure}[h!]
 \centering
\includegraphics{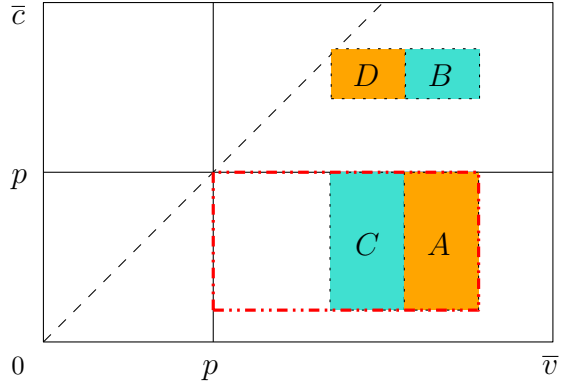}
\caption{Revenue decreasing redistribution for $U_G$ region}
\label{fig:ug}
\end{figure}

\end{proof}

\begin{figure}[h!]
\centering
\includegraphics[scale=1.1]{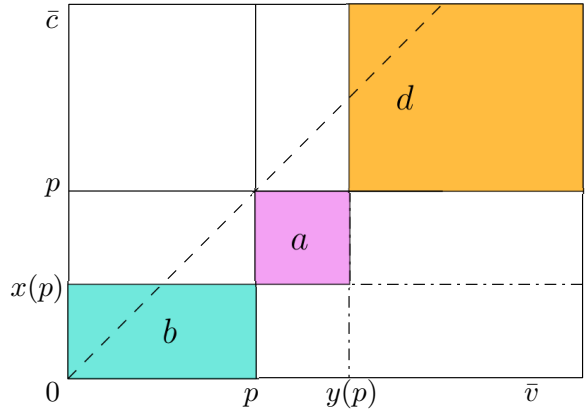}
\caption{Worst distribution when $\ell_p>0$}
\label{fig:worst2b}
\end{figure}

When $\ell_p>0$, for consistency with the marginals and the lemmas \ref{lem1} and \ref{lem2}, we will have regions with mass $b$ and $d$ as shown in Figure \ref{fig:worst2b}.

Here, 
\begin{align}
 \label{eq:const} a&=  \int_{0}^p g(c)~dc- \int_{0}^p f(v)~dv, \\
 \nonumber b&= \int_{0}^{x(p)} g(c)~dc = \int_{0}^p f(v)~dv, \\
 \nonumber d&= \int_{p}^{\overline{c}}g(c)~dc = \int_{y(p)}^{\overline{v}} f(v)~dv.
\end{align}
Note that here, $x(p) < p.$ \footnote{For $\ell_p< 0,$ $x(p) > p.$} For $p=0,$ by definition, $x(p)=0$ and the efficiency gains is zero. 

From the analysis above, depending upon $\ell_p$, we will have either of the two distributions as worst distribution. When $\ell_p \le 0$, the efficiency gains is zero. %As a result, instead of looking at all deterministic posted prices, we could look at just maximisation of efficiency gains over $p$ satisfying $x(p) \le p.$
Whenever the equations (\ref{eq:const}) are satisfied and $\ell_p>0$,  irrespective of how the mass is distributed within three rectangles, the efficiency gains are same, and given by 
\begin{equation*}
    \int_{p}^{y(p)} v f(v) ~dv- \int_{x(p)}^p c g(c) ~dc,
\end{equation*}
 where
  \begin{align*}
 \int_{0}^{x(p)} g(c)~dc &= \int_{0}^p f(v)~dv, \\
 \int_{y(p)}^{\overline{v}} f(v)~dv &=\int_{p}^{\overline{c}}g(c)~dc, \\
 x(p)&\le p.
 \end{align*}
 This gives us the first proposition of our paper.
 \begin{prop}\label{prop:1}
 The optimal robust mechanism within the class of deterministic posted price mechanism is mechanism with posted price $p^*$, where $p^*$ is solution to \begin{align*}
     \arg\max_{p \in \mathbb{R}_+} ~~ \int_{p}^{y(p)} v f(v) ~dv- \int_{x(p)}^p c g(c) ~dc,
 \end{align*}
% \footnote{Even if allow for randomised posted price mechanism, the optimal in the above sense is achieved by non-random posted price because of existence of a particular worst joint distribution at $p^*$ is approximately optimum. We explain this later in greater detail. }
where $G(x(p))=F(p) , F(y(p)) =G(p),$ and $x(p) \le p.$
 
 \end{prop}
 
Let $A= \max_p  \int_{p}^{y(p)} v f(v) ~dv- \int_{x(p)}^p c g(c) ~dc$. This represents the robust efficiency gains corresponding to optimal robust mechanism, and we will use it later. %- referred to as {\em robust gains from trade}.
 
Proposition \ref{prop:1} characterized that the optimal robust mechanism in the class of deterministic mechanisms. Our main result will show that this mechanism is also the optimal robust mechanism if we allowed for random mechanisms.

\subsection{Results on optimal robust mechanism}\label{optResult}
In this section, we show that the mechanism $M^{p^*}$ discussed in Proposition \ref{prop:1} is in fact the optimal robust
mechanism. We use the observation that for a collection of consistent joint distributions $\mathcal{F}(p^*)$, mechanism $M^{p^*}$ is the optimal mechanism and as a result, $M^{p^*} $ is optimal robust mechanism.  \footnote{For all $\epsilon$, the efficiency gains of mechanism, $M$ from the associated worst distribution cannot exceed the one corresponding to $M^{p^*}$ by an amount more than $\epsilon.$ } 

\noindent 
%We consider the following min-max problem:
 %$  \inf_{h \in \mathcal{H }^d(f,g)} \sup_{p \in R_+}  \int_{v>p} \int_{c<p} ~(v-c)h(v, c)~ dc ~dv \textrm{ and}$  show that it equals $\max_{{p \in R_+}} \int_{p}^{y(p)} v f(v) ~dv- \int_{x(p)}^p c g(c) ~dc \equiv A.$

 %\begin{prop} \label{thm:equiv1}
 %The value of max-min and min-max of gains from trade are equal in the class of deterministic posted price mechanisms.  
% $$ \inf_{h \in \mathcal{H }^d(f,g)}  \sup_{p \in R_+} \int_{v>p} \int_{c<p} ~(v-c)h(v, c)~ dc ~dv = \sup_{p \in R_+}  \inf_{h \in \mathcal{H }^d(f,g)} \int_{v>p} \int_{c<p} ~(v-c)h(v, c)~ dc ~dv$$
 %\end{prop}
  
%The proposition uses the fact that there exists a joint distribution $h(v,c)$ where gains corresponding to $p^*$ can't be less than the gains from trade corresponding 

\begin{figure}[h]
     \centering
     \begin{subfigure}[b]{0.47\textwidth}
         \centering
        \includegraphics{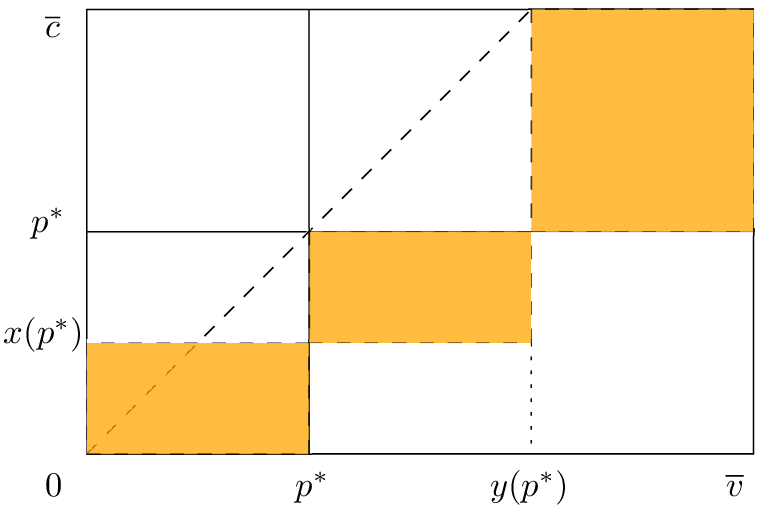}
         \caption{Coarse distribution}        
          \label{fig:coarse}
     \end{subfigure}
     \hfill
     \begin{subfigure}[b]{0.47\textwidth}
         \centering
         \includegraphics{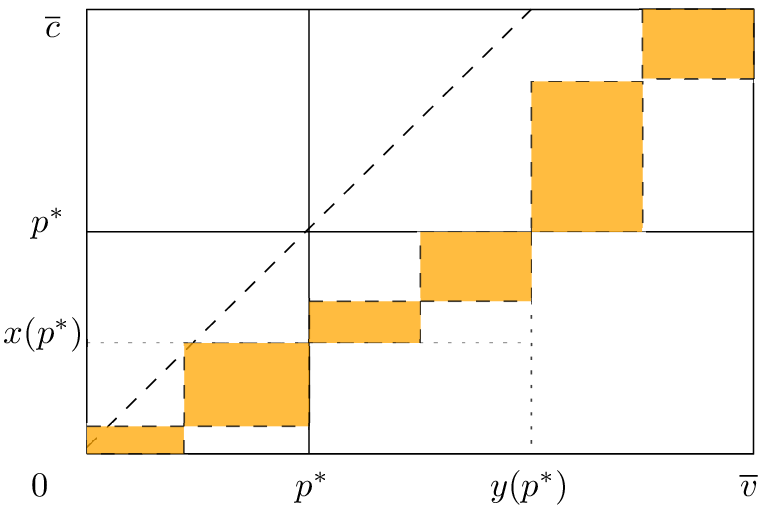}
         \caption{Fine distribution}     
          \label{fig:fine}
     \end{subfigure}
        \caption{Joint distributions in collection $\mathcal{F}(p)$}
        \label{fig:F}
\end{figure}

For posted price mechanism $M^{p^*}$, there is a worst distribution of the form that we found in Sub-section~\ref{worst}. In that form, we have three rectangles with positive mass. Consider the collection of consistent joint distributions $\mathcal{F}(p^*)$ having finer mass distribution than the obtained worst distribution. For illustration, consider Figure \ref{fig:F}. Three segments of buyer's valuation further divided into two equal parts. Given the six segments for buyer's valuation, we will have six segments of valuation of seller ensuring consistency in marginal distribution. Now, we will have six smaller rectangles with mass. Such a finer mass distribution ensures that efficiency gains by mechanism, $M^{p^*}$ for these joint distributions is same as the guaranteed efficiency gains associated with mechanism as these finer distributions are of the form of worst distribution associated with $M^{p^*}.$

%In particular, the efficiency gains from deterministic posted price mechanism, $M^{p^*}$ is $A$ for all $h(v,c) \in \mathcal{F}(p^*)$.

Consider any posted price mechanism $M$ with distribution function $G(p)$ over prices. The efficiency gains from $M$ for a finer distribution can possibly be greater than $A$. But as we make the distributions finer and finer, the efficiency gains converges to the robust efficiency gains from $M$.\footnote{The limiting distribution will be such that entire mass is on line, given by equation $F(v)=G(c)$. } As a result, the efficiency gains from the posted price mechanism is convex combination of robust efficiency gains of mechanisms, $M^p$. Since $M^{p^*}$ is the optimal robust mechanism in the class of deterministic posted price mechanisms $p^*$, the robust efficiency gains corresponding to $M^{p^*}$ will be greater than those corresponding to posted price mechanism. It follows that $M^{p^*}$ is optimal robust mechanism in $\mathcal{M}_P.$

\begin{theorem} \label{theo:rob}
The posted price mechanism $M^{p^*}$ is an optimal robust mechanism.
\end{theorem}

The theorem implies that if we are interested in optimal robust mechanisms in bilateral trading setting, we can restrict ourselves to a very simple class of mechanism--deterministic posted price mechanism.

We have considered the max-min problem till now, where designer wants to design an optimal robust mechanism and the efficiency gains associated with optimal robust mechanism gives the lower bound on the efficiency gains that can be realised if the designer optimally chooses the mechanism. The alternate lower bound for efficiency gains will be the one in which there is uncertainty about the true joint probability density but after the realised value of joint probability density, the designer can optimally choose mechanism. We would expect that the efficiency gains in such a situation will be higher as the designer gets to choose mechanism after realisation of true joint probability density in comparison to choosing the mechanism before realisation in case of max-min problem. However, as we see in the following theorem, these lower bounds actually coincide.

%The result in proposition \ref{thm:equiv1} can be further generalised from deterministic posted price mechanism to posted price mechanisms. It follows from Theorem \ref{theo:rob} and the fact that inf- sup problem over deterministic posted price mechanism is same as that with posted price mechanism as distribution enters linearly in objective function. 

\begin{theorem}\label{theo:equiv1}
The values of max-min and min-max of efficiency gains are equal for the class of DSIC mechanisms with BB and IR, \footnote{The equality holds even for more general setting of BIC mechanisms satisfying BB and interim individual rationality.} i.e.,
\begin{multline}
    \inf_{\hat{h} \in \mathcal{H}^d}  \sup_{ (q,t)\in \mathcal{M}_P}\int_{v} \int_{c} (v-c) \hat{h}(v,c) ~q(v,c)~dc ~dv = \sup_{(q,t) \in \mathcal{M}_P}  \inf_{\hat{h} \in \mathcal{H}^d}\int_{v} \int_{c} (v-c) \hat{h}(v,c) ~q(v,c)~dc ~dv.
\end{multline}
\end{theorem}

The above theorem holds because restricted to collection $\mathcal{F}(p^*),$ $p^*$ maximises efficiency gains. Restricted to these joint densities, had we chosen the mechanism optimally for true joint probability density, the guaranteed efficiency gains is $A= \textsc{Eff}(M^{p^*})$. Combined with the fact that min-max is not less than max-min in any problem, we get the value of max-min and min-max of efficiency gains equal. 

%Expansion to all the consistent joint distributions could reduce these guaranteed gains from trade, but since $M^{p^*}$ is optimal robust mechanism, the guaranteed efficiency gains would be $A$, which equals robust efficiency gains in $\mathcal{M}_P$

\section{Concluding remarks}

We studied distributionally robust mechanism design in bilateral trade setting. We focused on max-min version of robustness for joint distribution and found that there is no loss of generality in restricting attention to DSIC mechanisms. This allows us to use characterisation result for implementable mechanisms in the class of block mechanisms in \cite{Hagerty1987} and just focus on posted price mechanisms. 

We use the idea of redistribution of mass to characterise the worst distribution for deterministic posted price. For all deterministic posted price mechanism, maximally correlated distribution gives the worst expected gains from the trade. This feature is critical in proving optimality of deterministic posted price, and equivalence between max-min and min-max exercise.

Our analysis can be extended to an auction setting. \cite{he2022correlation} have considered this setting in auction environment but they focused on asymptotically optimal auction. Unlike the bilateral trade setting, in an auction setting, the worst distribution depends on the auction. This makes it more challenging to find the worst distribution.

\section*{Acknowledgement}
I am grateful to my supervisor, Debasis Mishra for the guidance. I appreciate Jeffrey Mensch and Hemant Mishra for their suggestions. I would also like to thank Aditya Vikram, Arunava Sen, Li Jiangtao, and seminar participants at $32^\textrm{nd}$ Stony Brook International Conference on Game
Theory and Royal Economic Society Conference (2021) for useful comments.
\bibliographystyle{ecta}
\bibliography{order_bilateral2}

\section{Appendix} \label{Appendix}
%\subsection{Proofs of Section XXX}

\subsection{Proof of Lemma \ref{lem:bb}}
 
\noindent We start by defining the following functions:
\begin{enumerate}
\item $\tilde{v}(c)= \inf\left[\left\{v: \lim_{r \rightarrow \infty}r^2 \int \limits_{c- \frac{1}{r}}^{c} \int \limits_{v}^{v + \frac{1}{r}} q(x,y)~dx~dy   >0 \right\} \cup \{1\}\right]$
\item
$\tilde{c}(v)= \sup \left[ \left\{c: \lim_{r \rightarrow \infty}r^2\int \limits_{c- \frac{1}{r}}^{c} \int \limits_{v}^{v + \frac{1}{r}} q(x,y)~dx~dy   >0 \right\} \cup \{ 0\}\right]$
\end{enumerate}
We define $\tilde{\Theta}=\{ (v,c): v > \tilde{v}(c), (v,c) \in \Theta \}$. Consider arbitrary $(v,c)\in \Theta$. By definition of $\tilde{v}(c)$, $\tilde{c}(v)$ and $(q^n, t^n_b, t^n_s)$, it follows that if $(v, c) \in \Theta - \tilde{\Theta},$ we have $q^n\left(v , c \right)=0 $ and $t^n_b(v,c)= t^n_s(v,c)=0$.\footnote{Note that $v \le \tilde{v}(c)$ iff $c \ge \tilde{c}(v).$}  Thus, we have budget balancedness for such values. 

Now we just need to show $\eta$-budget balancedness for $(v,c) \in \tilde{\Theta}$. Formally, we show that as $n \rightarrow \infty$, we have $|t^n_b(v,c)- t^n_s(v,c)| \rightarrow |t_b(v,c)- t_s(v,c)|.$ We start by finding relation between $t_b$ and $t_b^n.$

By equation (\ref{eq:payoff1}) and definition of $u_b)$ in \eqref{eq:IC}, we get
\begin{align}
\label{eq:myer1} \int \limits_{c-\frac{1}{n}}^c t_b(x,y)~dy & = \int \limits_{c- \frac{1}{n}}^c t_b(0,y)~dy + \int \limits_{c-\frac{1}{n}}^c x ~q(x, y)~dy - \int \limits_{c-\frac{1}{n}}^c \left[\int_0^x q(r, y) ~dr\right]~dy, ~~\forall x.
\end{align}
Integrating over $x$ from $v$ to $v+\frac{1}{n}$, we have 
\begin{multline}\label{eq:payoff2}
  \int \limits_{c-\frac{1}{n}}^c \int \limits_{v}^{v+ \frac{1}{n}} t_b(x,y)~dx~dy  = \dfrac{1}{n} \int \limits_{c- \frac{1}{n}}^c   t_b(0,y)~dy  + \int \limits_{c-\frac{1}{n}}^c \int \limits_{v}^{v+ \frac{1}{n}}x ~q(x, y)~dx~dy \\  -\int \limits_{c-\frac{1}{n}}^c\int \limits_{v}^{v+ \frac{1}{n}} \left[\int_0^x q(r, y) ~dr\right]~dx~dy.
 \end{multline}
 
We simplify the second term in right hand side of above equation using integration by parts. It gives
 \begin{align}
\nonumber  &\int \limits_{v}^{v+ \frac{1}{n}}x ~q(x, y)~dx \\
 \nonumber &\hspace{0.5cm} = \left[ x ~\int_0^x q(r, y)~dr \right]_{v}^{v+ \frac{1}{n}}- \int \limits_{v}^{v+ \frac{1}{n}} \int_0^x q(r, y) ~dr~dx\\
\nonumber & \hspace{0.5cm} = \left( v+ \frac{1}{n}\right) ~\int_0^{v+ \frac{1}{n} } q(r, y)~dr-  v ~\int_0^{v } q(r, y)~dr  - \int \limits_{v}^{v+ \frac{1}{n}} \int_0^x q(r, y) ~dr~dx\\
\label{eq:bypart} &\hspace{0.5cm} =v ~\int_v^{v+ \frac{1}{n} } q(r, y)~dr + \frac{1}{n}\int_0^{v+ \frac{1}{n} } q(r, y)~dr - \int \limits_{v}^{v+ \frac{1}{n}}\int_0^x q(r, y) ~dr~dx.
\end{align}
Substituting (\ref{eq:bypart}) in (\ref{eq:payoff2}), and multiplying the resultant equation by $n^2$, we get
\begin{align*}
 \nonumber & n^2\int \limits_{c-\frac{1}{n}}^c \int \limits_{v}^{v+ \frac{1}{n}} t_b(x,y)~dx~dy  \\
 &\hspace{0.5cm} = n \int \limits_{c- \frac{1}{n}}^c t_b(0,y)~dy +v  \left[ n^2 \int \limits_{c-\frac{1}{n}}^c \int \limits_{v}^{v+ \frac{1}{n}}~q(x, y)~dx~dy \right]+  n\left[ \int \limits_{c-\frac{1}{n}}^c \int \limits_{v}^{v+ \frac{1}{n}}~q(x, y)~dx~dy\right]\nonumber \\
   & \hspace{1.5cm} - 2n^2\int \limits_{c-\frac{1}{n}}^c\int \limits_{v}^{v+ \frac{1}{n}} \left[\int_0^x q(r, y) ~dr\right]~dx~dy\\
 &\hspace{0.5cm} =n \int \limits_{c- \frac{1}{n}}^c t_b(0,y)~dy + t^n_b(v,c) - t^n_b\left(v-\frac{1}{n}, c\right) + v q^n\left(v-\frac{1}{n}, c\right)\nonumber\\
   &\hspace{1.5cm} +n \int \limits_{c-\frac{1}{n}}^c \int \limits_{v}^{v+ \frac{1}{n}}~q(x, y)~dx~dy-
 2n^2\int \limits_{c-\frac{1}{n}}^c\int \limits_{v}^{v+ \frac{1}{n}} \left[\int_0^x q(r, y) ~dr\right]~dx~dy.
\end{align*}

The last inequality follows from definitions of $q^n$ and $t^n$ in \eqref{eq:all} and \eqref{eq:payb}, respectively.

Iteratively using relation between $(q^n,t^n_b)$ of consecutive blocks from equation (\ref{eq:payb}), we get
\begin{align}
  \label{eq:3} &n^2\int \limits_{c-\frac{1}{n}}^c \int \limits_{v}^{v+ \frac{1}{n}} t_b(x,y)~dx~dy  \\
&\hspace{0.5cm} =  n\int \limits_{c- \frac{1}{n}}^c   t_b(0,y)~dy+ t^n_b(v,c)  + \sum_{i=1}^{k-1} \frac{1}{n} q^n\left(v- \dfrac{i}{n}, c\right)- t^n_b\left( v-\frac{k}{n}, c \right) \nonumber \\
\nonumber &\hspace{1.5cm} + \left(v - \frac{k-1}{n} \right) q^n\left(v-\frac{k}{n}, c \right) +n \int \limits_{c-\frac{1}{n}}^c \int \limits_{v}^{v+ \frac{1}{n}}~q(x, y)~dx~dy  \\
 &\hspace{2.5cm} 
 - 2n^2\int \limits_{c-\frac{1}{n}}^c\int \limits_{v}^{v+ \frac{1}{n}} \int_0^x q(r, y) ~dr~dx~dy,
 \end{align}
where $k(n)=\min \left\{k: q^n\left(v- \dfrac{ k}{n}, c\right)=0 , k \in \mathbb{N} \right\}.$ \footnote{It is well defined function as $v> \tilde{v}$ and by definition of $q^n$, whenever $(v,c)$ lie on boundary of grid, $q^n(v,c)>0.$}\\

We will simplify the above expression by using following observations. By definition of $k$ and individual rationality of $q^n$, we have 
\begin{align}
   \label{eq:1}t^n_b \left( v- \dfrac{k}{n}, c\right)&=q^n\left( v- \dfrac{k}{n}, c\right)=0,\textrm{ and }\\
    \label{eq:2} \sum_{i=1}^{k-1} \dfrac{1}{n} q^n\left(v- \dfrac{i}{n}, c\right)&= n \int \limits_{c- \frac{1}{n}}^c \int \limits_{v- \frac{k-1}{n}}^v q(x,y)~dx ~dy
\end{align}

Using \eqref{eq:1} and \eqref{eq:2} in \eqref{eq:3} gives
\begin{align}
 \nonumber n^2\int \limits_{c-\frac{1}{n}}^c& \int \limits_{v}^{v+ \frac{1}{n}} t_b(x,y)~dx~dy  \\
 \nonumber & = n \int \limits_{c- \frac{1}{n}}^c t_b(0,y)dy + t^n_b(v,c) + n \int \limits_{c- \frac{1}{n}}^c \int \limits_{v- \frac{k-1}{n}}^v q(x,y)~dx ~dy+  n \int \limits_{c-\frac{1}{n}}^c \int \limits_{v}^{v+ \frac{1}{n}}~q(x, y)~dx~dy\\
\label{eq:main}&~~~- 2n^2\int \limits_{c-\frac{1}{n}}^c\int \limits_{v}^{v+ \frac{1}{n}} \int \limits_{\tilde{v}(c)}^x q(r, y) ~dr~dx~dy
 -  2n^2\int \limits_{c-\frac{1}{n}}^c\int \limits_{v}^{v+ \frac{1}{n}} \int^{\tilde{v}(c)}_0q(r, y) ~dr~dx~dy. 
 \end{align}
Thus, we have
 \begin{align}
  \label{eq:6}n^2\int \limits_{c-\frac{1}{n}}^c\int \limits_{v}^{v+ \frac{1}{n}} t_b(x,y)~dx~dy &=   t_b^n(v,c) + A_b(n,v,c)+ B_b(n,v,c)+ C_b(n,v,c),
  \end{align}
where
 \begin{align}
  \nonumber A_b(n,v,c) =& ~n \int \limits_{c- \frac{1}{n}}^c \int \limits_{v- \frac{k-1}{n}}^v q(x,y)~dx ~dy -  n \int \limits_{c-\frac{1}{n}}^c \int \limits_{\tilde{v}(c)}^{v}~q(x, y)~dx~dy,\\
 \nonumber B_b(n,v,c)= & ~n \int \limits_{c-\frac{1}{n}}^c \int^{v+ \frac{1}{n}}_{\tilde{v}(c)}~q(x, y)~dx~dy - 2n^2\int \limits_{c-\frac{1}{n}}^c\int \limits_{v}^{v+ \frac{1}{n}} \int \limits_{\tilde{v}(c)}^x q(r, y) ~dr~dx~dy, \\
  \label{eq:7}C_b(n,v,c) =&  -  2n^2\int \limits_{c-\frac{1}{n}}^c\int \limits_{v}^{v+ \frac{1}{n}} \int^{\tilde{v}(c)}_0q(r, y) ~dr~dx~dy +n \int \limits_{c- \frac{1}{n}}^c  t_b(0,y)~dy.
  \end{align}
From boundedness of $q$, we have
\begin{align}
 \label{eq:4}\lim_{n \rightarrow \infty }  A_b(n,v,c) =\lim_{n \rightarrow \infty }  B_b(n,v,c)=0.
\end{align}
Using M-IIR of $(q, t_b, t_s) \in \mathcal{M}_B$, we have
\begin{align}
    \int \limits_{c- \frac{1}{n}}^c  t_b(0,y)~dy=0.
\end{align}
Using this along with boundedness of $q$ in equation \eqref{eq:7}, we get
\begin{align}
   \label{eq:5} \lim_{n \rightarrow \infty }  C_b(n,v,c)&=0
\end{align}
Using \eqref{eq:4} and \eqref{eq:5} in \eqref{eq:6} we get that for $v> \tilde{v}(c)$,
\begin{align}
 \label{eq:limit1}\lim_{n \rightarrow \infty }n^2\int \limits_{c-\frac{1}{n}}^c &\int \limits_{v}^{v+ \frac{1}{n}} t_b(x,y)~dx~dy
 = \lim_{n\rightarrow \infty} t_b^n(v,c).
 \end{align}
Analogously, for each $(v,c)$ such that $c< \tilde{c}(v),$ we have
 \begin{align}
\label{eq:limit2}	\lim_{n \rightarrow \infty } &n^2  \int \limits_{c-\frac{1}{n}}^c \int \limits_{v}^{v+ \frac{1}{n}} t_s(x,y)~dx~dy  = \lim_{n \rightarrow \infty}t^n_s(v, c).
\end{align}
Subtracting \eqref{eq:limit2} from \eqref{eq:limit1}, for $v> \tilde{v}(c)$ and $c< \tilde{c}(v)$, we get 
\begin{align}
\label{eq:24} \lim_{n \rightarrow \infty}  &t^n_b(v,c)- t^n_s(v, c)=\lim_{n \rightarrow \infty }n^2\int \limits_{c-\frac{1}{n}}^c\int \limits_{v}^{v+ \frac{1}{n}}\left[ t_b(x,y)- t_s(x,y) \right]~ dx ~dy=0.
\end{align}
The last equality follows from budget balancedness of $(q,t_b,t_s).$\\

We have thus shown that for $(v,c) \in \tilde{\Theta},$ the $\eta$-budget balancedness holds for a large enough $n$. From budget balancedness for $(v,c) \in \Theta- \tilde{\Theta}$ and $\eta$-budget balancedness for $(v,c)\in \tilde{\Theta}$, we get that for all $ (v,c)\in \Theta$, 
	\begin{align*}
	    \lim_{n \rightarrow \infty } | t^n_b(v,c)- t^n_s(v, c)| =  0.
	\end{align*}

\subsection{Proof of Lemma \ref{lem:conv2}}
\noindent {\sc Step 1.} We define an allocation rule $ \displaystyle \hat{q}^r (v,c):= r^2 \int \limits_{B_{k',\ell'}(r)} q(x,y)~dx~dy,$ where $(v,c) \in B_{k',\ell'}(r).$ \\

Recall the following definitions:
\begin{enumerate}
\item $\tilde{v}(c)= \inf \left[\left\{v: \lim_{r \rightarrow \infty}r^2 \int \limits_{c- \frac{1}{r}}^{c} \int \limits_{v}^{v + \frac{1}{r}} q(x,y)~dx~dy   >0 \right\}\cup \{1\}\right]$
\item
$\tilde{c}(v)= \sup \left[ \left\{c: \lim_{r \rightarrow \infty}r^2\int \limits_{c- \frac{1}{r}}^{c} \int \limits_{v}^{v + \frac{1}{r}} q(x,y)~dx~dy   >0 \right\}\cup \{0\}\right]$
\end{enumerate}

We show the convergence between probability of allocation for the given mechanism, $q^n(v,c)$ and  $\hat{q}^n(v,c)$.\\

\noindent Let $B= \{ (v,c): v= \tilde{v}(c) \textrm{ or }c= \tilde{c}(v) \}$ and $\mathcal{D}(wm)=\{ B_{k',\ell'}(wm) : (v,c) \in B_{k',\ell'}(wm) \cap C \}.$\\\\
Notice that for a given $r$, $\displaystyle q^r(v, c ) \ne  r^2 \int \limits_{B_{k, \ell}(r)} q(x,y)~dx~dy = \hat{q}^r(v,c)$ for squares, $B_{k',\ell'}(r)$ containing a point in $B$. In particular, restricted to $B_{k, \ell}(w)$, the values differ over $\mathcal{D}(wm)$ only.  It follows directly from the definition of $q^{wm}(v,c).$ We show this below.\\

Suppose for contradiction that there exists $(v,c)\in B_{k',\ell'}(wm) \notin \mathcal{D}(wm)$ and $\hat{q}^{wm}(v,c)\ne q^{wm}(v,c)$.
Note that $\hat{q}^{wm}(v,c)\ne q^{wm}(v,c)$ implies that $\hat{q}^{wm}(v,c)>0 = q^{wm}(v,c)$.
	Since $\hat{q}^{wm}(v,c)>0 $, we have $x> \tilde{v}(y)$  and $y< \tilde{c}(x)$ for all $(x,y) \in B_{k',\ell'}(wm).$\footnote{ The inequalities are strict as $B_{k',\ell'}(wm) \notin \mathcal{D}(wm)$.} %This implies  $\min \{x: x \in I_{k'}(nm)\}> \max_{y\in I_{l'}(nm)} \tilde{v}(y)$  and $\max \{ y \in I_{l'}(nm) \}< \min_{x \in I_{k'}(nm)} \tilde{c}(x)$.
	Also, by definition of $q^{wm}$, $ q^{wm}(v,c)=0$ implies that  $ \hat{q}^{wm}(v-\frac{1}{wm},c)=0$ or $\hat{q}^{wm}(v, c +\frac{1}{wm})=0$. Combining with the fact $\hat{q}^{wm}(v,c)>0$, we get $ \tilde{v}(y) \le x$ or $ \tilde{c}(x) \ge y,$ for some $(x,y) \in B_{k',\ell'}(wm),$ which is a contradiction.\\
	
Using the above fact, for all $B_{k, \ell(w)}$ we have
\begin{align*}
 \bigintssss \limits_{B_{k, \ell}(w)} \hat{q}^{wm}(v,c) ~d\widehat{H}(v,c)& -\bigintssss \limits_{B_{k, \ell}(w)} q^{wm}(v,c) ~d\widehat{H}(v,c)\\
 &= \sum_{ B_{i,j}(wm)  \subseteq B_{k, \ell}(w) } \left[ \bigintssss \limits_{B_{i,j}(wm)} \hat{q}^{wm}(v,c) ~d\widehat{H}(v,c) -\bigintssss \limits_{B_{i,j}(wm)} q^{wm}(v,c) ~d\widehat{H}(v,c)\right]\\
 &\le \sum_{D \in \mathcal{D}(wm) } \int \limits_{D} \hat{q}^{wm}(v,c) ~d\widehat{H}(v,c) \\
&\le \sum_{D \in \mathcal{D}(wm) } \widehat{H}(D).
\end{align*}
The first equality comes from partioning each block into $m$ blocks. The first inequality follows from the fact that for all $ (v,c), ~\hat{q}^{wm}(v,c) \ge q^{wm}(v,c)$ and that $q^{wm}$ is non-negative. The last inequality follows from the fact that $\hat{q}^{wm}(v,c) \le 1$ by definition. 
Thus, for all $B_{k, \ell(w)}$, as $m \rightarrow \infty, $ we have
\begin{multline*}
 \lim_{m \rightarrow \infty } \left[ \bigintssss \limits_{B_{k, \ell}(w)} \hat{q}^{wm}(v,c) ~d\widehat{H}(v,c) -\bigintssss \limits_{B_{k, \ell}(w)} q^{wm}(v,c) ~d\widehat{H}(v,c) \right] \\ \le \lim_{m \rightarrow \infty } \sum_{D \in \mathcal{D}(wm)} \widehat{H}(D)=\widehat{H}(B)=0.
 \end{multline*}
 Alternatively, for all $B_{k, \ell(w)}$, as $m \rightarrow \infty,$ we have
 \begin{align*}
 \bigintssss \limits_{B_{k, \ell}(w)} &q^{wm}(v,c) ~d\widehat{H}(v,c) \rightarrow
 \bigintssss \limits_{B_{k, \ell}(w)} \hat{q}^{wm}(v,c) ~d\widehat{H}(v,c).
 \end{align*}\\

\noindent {\sc Step 2. }We show the convergence between probability of allocation for the given mechanism $q(v,c)$ and  $\hat{q}^n(v,c)$, which would later be used to prove Lemma \ref{lem:got}.\\

To simplify notations, we consider two measures on the Borel sigma algebra: Lebesgue measure, $\lambda$ and given probability measure, $\widehat{H}.$ We use the concept of simple functions to show convergence in probabilities. 

Consider a rectangle, $I$. We can divide it into $\alpha^2$ equal sized blocks by cutting each side into $\alpha$ equal parts. We define $\mathcal{S}(I,\alpha)$ as the collection of these $\alpha^2$ equal sized blocks. 

Let $\psi : B_{k, \ell}(w) \rightarrow \mathbb{R}$ be a simple function defined as $\psi(v,c)= \sum_i^r a_i \mathcal{X}_{A_i}(x)$ where $\mathcal{X}_{A_i}$ is indicator function, ${A_i}$ are disjoint measurable sets and $\psi(v,c) \ge q(v,c), $ for all $(v,c). $ Also, we have $\cup_{i=1}^r A_i= B_{k, \ell}(w).$

Fix collection of measurable sets, $\mathcal{A}=\{ A_1, A_2 , \dots, A_r \}$. Since $A_i$ is a measurable set, for every $\epsilon>0$, there exists collection of blocks (squares) (for large enough $m_i$) $\{ S^i_j: j \in \mathbb{N}\} \subset \mathcal{S}(B_{k, \ell}(w),~m_i)$ such that
\begin{enumerate}[label=(\roman*)]
\item $ \bigcup_j S^i_{j} \supseteq A_i,$
\item $\lambda\left(\bigcup_j S^i_{j}\right)< \lambda(A_i) + \epsilon,$   and
 \item $\widehat{H}\left(\bigcup_j S^i_{j} \right)< \widehat{H}(A_i) + \epsilon.$
 \end{enumerate}
Choose $m= \max_i m_i.$ This ensures that above conditions hold simultaneously for all $A_i.$\\

\noindent Note that
 \begin{align}
 \nonumber \lambda\left(\left(\bigcup_j S^i_j \right)\cap \left(\bigcup_{t \ne i} A_t \right)\right)&=  \lambda \left(\bigcup_j S^i_{j}\right)  + \lambda  \left(\bigcup_{t \ne i} A_t\right)- \lambda \left(\left(\bigcup_j S^i_{j}\right) \bigcup \left(\bigcup_{t \ne i} A_t\right)\right) \\
 \nonumber&< \lambda(A_i) + \epsilon + \sum_{t \ne i}\lambda( A_t)- \lambda(B_{k, \ell}(w))\\
 \label{eq:10}&=\epsilon.
 \end{align}
The first equality follows from additivity of measure $\lambda$ for disjoint sets. The strict inequality follows from definition of $S_j^i$ and sub-additivity of measure $\lambda.$ The last equality follows from the fact that $\cup_{i=1}^r A_i= B_{k,l}(w).$ \\\\
Now consider arbitrary collection $\mathcal{I} \subseteq \mathcal{S}(B_{k, \ell}(w) ,~m)$ such that $\lambda(\mathcal{I}~ \cap~ A_i ) \ne \emptyset$, and $\lambda \left(\mathcal{I} ~\cap~ A_{i'})\right ) \ne \emptyset,$ %since $I \subset \cup_j S_j^i$, we get % 
we have
\begin{align}
  \label{eq:11}\lambda \left( \mathcal{I} \cap A_i\right)&< \epsilon \textrm{ and}\\
  \label{eq:12}\lambda \left( \mathcal{I} \cap A_{i'}\right)&< \epsilon.
\end{align}
This follows from the fact that $\mathcal{I} \subseteq \{ S_j^i\}$ and $\mathcal{I} \subseteq \{ S_j^{i'}\}$ and equation \eqref{eq:10}. This is explained below.

Note that
\begin{align*}
  \lambda(\mathcal{I} \cap A_{i'})&\le\lambda\left(\mathcal{I} \cap \left(\bigcup_{t \ne i} A_t\right)\right)\\
  &\le\lambda\left(\left(\bigcup_j S^i_j \right)\cap \left(\bigcup_{t \ne i} A_t \right)\right)\\
  &<\epsilon
\end{align*}
where last inequality follows from equation~\eqref{eq:10}.

Analogously, we get the expression for $A_{i}$ as presented in \eqref{eq:11}. We use use this equation in this proof later.

\noindent Now we show that $\bigintssss \limits_{B_{k, \ell}(w)} \hat{q}_{wm }(x,y) d\widehat{H}(x,y) \le \bigintssss \limits_{B_{k, \ell}(w)} \psi (x,y) d\widehat{H}(x,y)$ as $m \rightarrow \infty.$\\\\
We start by finding expected $\hat{q}$ over arbitrary $A_i.$
Consider arbitrary $I \in \mathcal{S}(B_{k, \ell}(w), ~m).$ For $(v,c) \in I $, we have \vspace{0.08in}
\begin{spreadlines}{1.2em}
 \begin{align}
\nonumber \hat{q}^{wm}(v,c)&=\dfrac{ \int \limits_{I} q(x,y) ~d\lambda(x,y)}{\lambda(I)} \\
\nonumber&\le   \sum_t a_t\dfrac{  \lambda(A_t \cap I)}{\lambda(I)} \\
 \label{eq:9}&\le   a_i\dfrac{ \lambda(A_i \cap I) }{\lambda(I)}+ \max_t{a_t} \sum_{t\ne i}  \dfrac{\lambda (A_t \cap I) }{\lambda(I)}.
\end{align}
\end{spreadlines}
The equality follows from the definition of $\hat{q}_n(v,c)$. The first inequality follows from definition of function of $\psi$ and the last inequality follows from definition of max.
Thus,
\begin{align}
 \int \limits_{A_i} \hat{q}^{wm}(x,y) ~d\widehat{H}(x,y)&\le \sum_{I \in \mathcal{S}(B_{k, \ell}(w),~ m)} \left[ a_i \dfrac{\lambda(A_i \cap I)}{{\lambda(I)}} + \max_t{a_t} \sum_{t\ne i}  \dfrac{\lambda(A_t \cap I)}{\lambda(I)}\right] \widehat{H}(A_i \cap I).
 \end{align}

We partition the set $\mathcal{S}(B_{k, \ell}(w), ~m)$ on the basis of whether $I$ intersects measurable sets other than $A_i$ or not, with positive measure. Consider  $\mathcal{R}(A_i)= \{ I: \lambda (I \cap \cup_{t \ne i}A_t)=0 \}.$
For  $(v,c )\in I \in \mathcal{R}(A_i)$, by equation~\eqref{eq:9}, we get $\hat{q}^{wm}(v,c)\le a_i.$ This implies  
 \begin{align}
\label{eq:conv1}  \sum_{I \in \mathcal{R}(A_i) }\hat{q}^{wm}(x,y)\widehat{H}(A_i \cap I) &\le  a_i \sum_{I \in \mathcal{R}(A_i) }\widehat{H}(A_i \cap I).
\end{align}
For $(v,c) \in I \notin \mathcal{R}(A_i)$, we have
 \begin{align}
 \nonumber \sum_{ I \notin \mathcal{R}(A_i) }\hat{q}^{wm}(x,y)\widehat{H}(A_i \cap I) &\le \sum_{I \notin \mathcal{R}(A_i) }\left[ a_i \dfrac{\lambda(A_i \cap I)}{{\lambda(I)}} + \max_t{a_t} \sum_{t\ne i}  \dfrac{\lambda(A_t \cap I)}{{\lambda(I)}}\right] \widehat{H}(A_i \cap I) \\
   \nonumber &\le \sum_{I \notin \mathcal{R}(A_i) } \left[ a_i + \max_t (a_t - a_i) \right] \widehat{H}(A_i \cap I)\\
 \nonumber &= a_i \sum_{I \notin \mathcal{R}(A_i) }  \widehat{H}(A_i \cap I) + \max_t (a_t - a_i) \sum_{I \notin \mathcal{R}(A_i) }\widehat{H}(A_i \cap I)\\
 &=a_i \sum_{I \notin \mathcal{R}(A_i) }  \widehat{H}(A_i \cap I) + \max_t (a_t - a_i)\widehat{H}(A_i \cap ( \cup_{I \notin \mathcal{R}(A_i)} I))\\
\label{eq:conv2} &< a_i \sum_{I \notin \mathcal{R}(A_i) }  \widehat{H}(A_i \cap I) + \max_t (a_t - a_i) (r-1)\epsilon.
\end{align}
The last strict inequality uses equation~\eqref{eq:11} and the fact that we are considering $r$ number of measurable sets. 
%&\le \left[ a_i + \max_t (a_t - a_i) \right] H(A_i \cap ( \cup_{I}B_{k, \ell}))\\
  %& < a_i H(A_i \cap ( \cup_{I}B_{k, \ell})) + \max_t (a_t - a_i) \epsilon
Adding (\ref{eq:conv1}) and (\ref{eq:conv2}), we get that $\forall  i \in \{ 1, \dots, r\},$
 \begin{align}
     \nonumber \int \limits_{A_i} \hat{q}^{wm}(x,y) ~d\widehat{H}(x,y)&< a_i \widehat{H}(A_i)+ \max_t (a_t -a_i)( r-1) \epsilon.
\end{align}
 Summing over $i$, we get 
 \begin{align}
   \nonumber    \int \limits_{B_{k, \ell}(w)} \hat{q}^{wm}(x,y) ~d\widehat{H}(x,y)& < \int \limits_{B_{k, \ell}(w)} \psi(x,y)~d\widehat{H}(x,y)+\sum_i \max_t (a_t -a_i)(r-1) \epsilon.
 \end{align}
 As the above equation holds for all $\epsilon>0$, we get
 \begin{align}
      \int \limits_{B_{k, \ell}(w)} \hat{q}^{wm}(x,y) ~d\widehat{H}(x,y) &\le \int \limits_{B_{k, \ell}(w)}  \psi (x,y)~d\widehat{H}(x,y).
 \end{align}
 
Now we will show similar result for simple functions approaching $q$ from below.
Consider $\phi : B_{k, \ell}(w) \rightarrow \mathbb{R}$ be a simple function defined as $\phi(v,c)= \sum_i^r a_i \mathcal{X}_{A_i}(x),$ where $\mathcal{X}_{A_i}$ is indicator function, ${A_i}$ are disjoint measurable sets, and $\phi(v,c)\le  q(v,c), ~ \forall (v,c). $ Also, we have $\cup_{i=1}^r A_i= B_{k, \ell}(w).$\\

Analogously, we can show that $\bigintssss \limits_{B_{k, \ell}(w)} \hat{q}^{wm }(x,y) ~d\widehat{H}(x,y) \ge  \bigintssss \limits_{B_{k, \ell}(w)} \phi (x,y) ~d\widehat{H}(x,y)$ as $m \rightarrow \infty.$
Since $q(v,c)$ is integrable with respect to $H,$ we have
\begin{align}
\nonumber \inf_{\psi} \bigintssss \limits_{B_{k, \ell}(w)} \psi (x,y) ~d\widehat{H}(x,y)&=\sup_{\phi} \bigintssss \limits_{B_{k, \ell}(w)} \phi (x,y) ~d\widehat{H}(x,y).
\end{align}
Using the above fact with lower and upper bound for $ \bigintssss \limits_{B_{k, \ell}(w)} \hat{q}^{wm }(x,y) ~d\widehat{H}(x,y)$, we get that as $m \rightarrow \infty$, 
\begin{align}
  \nonumber \bigintssss \limits_{B_{k, \ell}(w)} \hat{q}^{wm }(x,y) ~d\widehat{H}(x,y) & \rightarrow  \bigintssss \limits_{B_{k, \ell}(w)} q(x,y) ~d\widehat{H}(x,y). 
  \end{align}
   {\sc Step 3.} We use the observations in {\sc Step 1} and {\sc Step 2} to prove lemma.
   
Combining the convergences in {\sc Step 1} and {\sc Step 2}, 
\begin{align}
  \label{eq:conv_p} \lim_{m \to \infty} \bigintssss \limits_{B_{k, \ell}(w)} q^{wm }(x,y) ~d\widehat{H}(x,y) & = \bigintssss \limits_{B_{k, \ell}(w)} q(x,y) ~d\widehat{H}(x,y).
  \end{align}
%Now, we would use the convergence in allocation probabilities to show convergence of efficiency gains.\\\\

Let $\Pi_1(\cdot)$ and $\Pi_2(\cdot)$ be the projection of a set on $x-$axis and $y-$axis, respectively.
Note that
 \begin{align}
\nonumber \bigg| ~~ \bigintssss \limits_{B_{k, \ell}(w)}& (x-y)~q^{wm}(x,y)~ d\widehat{H}(x,y) - \bigintssss \limits_{B_{k, \ell}(w)} (x-y)~q(x,y) ~d\widehat{H}(x,y)~~\bigg|\\
\nonumber&\le \left (\min \Pi_1(B_{k, \ell}(w)) - \inf \Pi_2(B_{k, \ell}(w))\right)  \bigg| ~~  \bigintssss \limits_{B_{k, \ell}(w)} q^{wm}(x,y) d\widehat{H}(x,y) - \bigintssss \limits_{B_{k, \ell}(w)} q(x,y) d\widehat{H}(x,y)~~ \bigg| \\
\label{eq:conv_v} &+ \dfrac{2}{w} \left[  \int \limits_{B_{k, \ell}(w)} \left(q^{wm}(x,y) + q(x,y) \right)d\widehat{H}(x,y)\right].
\end{align}
It follows from the fact that $ \Pi_1(\min B_{k, \ell}(w)) \le x< \Pi_1(\min B_{k, \ell}(w))+ \frac{1}{w}$ and $\inf \Pi_2(B_{k, \ell}(w)) < y\le \Pi_2( \inf B_{k, \ell}(w))+ \frac{1}{w}$.\\

Using equation (\ref{eq:conv_p}) and (\ref{eq:conv_v}), we get that as $m \rightarrow \infty$, we have
\begin{align}
 \nonumber \bigintssss \limits_{B_{k, \ell}(w)} (x-y)~&q^{wm}(x,y)~ d\widehat{H}(x,y)  \rightarrow   \bigintssss \limits_{B_{k, \ell}(w)} (x-y)~q(x,y) ~d\widehat{H}(x,y)+  \dfrac{2}{w} \left[  \int \limits_{B_{k, \ell}(w)}  q(x,y) ~d\widehat{H}(x,y)\right].
\end{align}

Combining Lemma~\ref{lem:ic}, Lemma~\ref{lem:bb}, Lemma~\ref{lem:got}, and Lemma~\ref{lem:dsic}, we get the following result:
for every $\epsilon>0, (q, t_b, t_s) \in \mathcal{M}_B, \widehat{H} \in \mathcal{H}, $ there exist $(q_0,t_0,t_0) \in \mathcal{M}_D$ such that $\textsc{Eff}(q,t_b, t_s)- \textsc{Eff}(q^n, t_b^n, t^n_s) \le \epsilon. $ This result is much stronger than the theorems stated in Section \ref{result}.

\subsection{Proof of observations in proof of Theorem \ref{theo:main2}}

{\bf Proof of Observation \ref{ob1}}

\noindent Consider arbitrary $(v_k, c_{\ell}).$
\noindent By equation (\ref{eq:payoff1}), we have
\begin{align}
\nonumber t_b(v_k, c_{\ell})&= v_k q(v_k, c_{\ell}) - \dfrac{1}{n}\sum_{i=0}^{k-1} q(v_i, c_{\ell}) + t_b(v_0,c_{\ell}),\\
\nonumber t_s(v_k, c_{\ell})&= c_{\ell} q(v_k, c_{\ell}) +\dfrac{1}{n} \sum_{i=l+1}^{n} q(v_k, c_i) + t_s(v_k, c_n).
\end{align}
Thus, subtracting the above two equations, we get
\begin{align}
\nonumber  t_b(v_k, c_{\ell})-t_s(v_k, c_{\ell})&=(v_k- c_{\ell}) q(v_k, c_{\ell}) -\dfrac{1}{n} \left[\sum_{i=0}^{k-1} q(v_i, c_{\ell})
+ \sum_{i=l+1}^{n} q(v_k, c_i) \right] + t_b(v_0, c_{\ell})- t_s(v_k, c_n).
\end{align}

\noindent {\bf Proof of Observation \ref{ob2}}

\noindent Fix a $v$ and $(c_{\ell-1},c_{\ell}]$. Define a joint density $\hat{h}^v$ as follows:
\begin{align*}
\hat{h}^v(x,y) =
\begin{cases}
n~f(v) & \textrm{if}~x=v, y \in (c_{\ell-1},c_{\ell}], \\
0 & \textrm{if}~x=v, y \notin (c_{\ell-1},c_{\ell}], \\
h(x,y) & \textrm{otherwise}.
\end{cases}
\end{align*}
\noindent By M-IIR of $(q, t_b, t_s) \in \mathcal{M}_{B},$ 
\begin{align*}
\int  \limits_{c_{\ell-1}}^{c_{\ell}}( t^n_b(v,y)- t^n_s(v,y))~dy&\le  \int \limits_{c_{\ell-1}}^{c_{\ell}} (v-y)q(v,y)~dy.
\end{align*}
The inequality holds for all $v$. By integrating over $v$ from $v_k $ to $v_{k+1}$, we get
\begin{align}
\label{eq:ir1}\int \limits_{v_k}^{v_{k+1}} \int \limits_{c_{\ell-1}}^{c_{\ell}}( t^n_b(x,y)- t^n_s(x,y))~dy~dx&\le \int \limits_{v_k}^{v_{k+1}} \int \limits_{c_{\ell-1}}^{c_{\ell}} (x-y)q(x,y)~dy~dx.
\end{align}
Analogously, by BB of $(q,t_b,t_s)\in \mathcal{M}_{B}$, we get
\begin{align}
\label{eq:ir2}\int \limits_{v_k}^{v_{k+1}} \int \limits_{c_{\ell-1}}^{c_{\ell}} (t^n_b(x,y)- t^n_s(x,y))~dy~dx&= 0.
\end{align}
Combining the inequalities (\ref{eq:ir1}) and (\ref{eq:ir2}), we get
\begin{align}
 \int \limits_{v_k}^{v_{k+1}}\int \limits_{c_{\ell-1}}^{c_{\ell}}(v-c)~q(x,y)~dy~dx&= 0.
\end{align}
For $k< \ell-1$, we have $(x-y)<0$ for all $(x, y)\in [v_k, v_{k+1})\times ( c_{\ell-1}, c_{\ell}]$. To satisfy the above inequality, we must have $q(v,c)=0$ a.e. 
Thus, for $k< \ell$, we have $q^n(v_k,c_{\ell})=0$ by definition of $q^n.$\\

\noindent
{\bf Proof of Observation \ref{obs3}}

% \noindent Intuitively, since $r_n(k', \ell')$ is defined iteratively and is increasing in surplus $t_b^n(v_{i'}, c_{j'})- t_s^n(v_{i'}, c_{j'})$, the maximum value of $r_n(k', \ell')$ is bounded above by case where when $t_b^n(v_{i'}, c_{j'})- t_s^n(v_{i'}, c_{j'})= \max_{i,j} t_b^n(v_{i}, c
% _{j})- t_s^n(v_{i}, c_{j})$ for all $i'>j'$.

We prove this observation by induction. For $k'=\ell'+1$, it holds true. Suppose the inequality holds for all $(i, j)$ where $i- j\le d$ and $d \in \{1,2, \dots, n-2\}$. We show that it holds for all $(k', \ell')$ where $k'- \ell'=d+1.$ Consider arbitrary $(k', \ell')$ with $k'-\ell'=d+1.$

 To simplify notation in the proof, let $\beta\coloneqq \max_{i,j} [t_b^n(v_i, c
_j)- t_s^n(v_i, c_j)]$. 
By the definition of $r_n$, we have
\begin{align}
\nonumber r_n(k', \ell')&\le \dfrac{n}{k'-\ell'} \left(\beta + \dfrac{1}{n}\sum_{i=\ell'+1}^{k'-1} (\overline{r}_n(k,i) + \overline{r}_n(i,\ell')) \right).
\end{align}
As the observation holds for $(i,j)$ where $i-j\le d$, and $\overline{r}_n(i,j)$ depends only on $i-j,$ we get
\begin{align}
\nonumber r_n(k', \ell')&\le \dfrac{n}{k'-\ell'} \left(\beta + \dfrac{2}{n}\sum_{i=2}^{k'-\ell'} \overline{r}_n(i, 1) \right)\\
\nonumber&=\dfrac{n}{k'-\ell'} \beta + \dfrac{2}{k'-\ell'} \sum_{i=2}^{k'-\ell'} \dfrac{i}{2}n\beta\\
\nonumber&=\left( \dfrac{n}{k'-\ell'} + \dfrac{1}{k'-\ell'} \sum_{i=2}^{k'-\ell'} ni \right)\beta \\
\nonumber&=\left( \dfrac{n}{k'-\ell'} + \dfrac{1}{k'-\ell'} \dfrac{ (2+ k'- \ell')(k'-\ell'-1)}{2}n \right)\beta\\
\nonumber&=\dfrac{k'-\ell'+1}{2}n\beta.
\end{align}

\subsection{  Proof of Theorem \ref{theo:rob}}

\noindent For $p^*$, we define a collection of consistent joint densities $F(p^*)$. It has distributions that are finer than the worst distribution for deterministic posted price mechanism $M^{p^*}$ depicted in Figure \ref{fig:worst2b}. In particular, the mass in each of the three rectangles is redistributed into some $n$ rectangles such that length of rectangles is same within each of the three rectangles. Formally,
\begin{multline*}
    \mathcal{F}(p^*)= \bigg\{ \hat{h}: \exists  ~n \in \mathbb{N} ~\& ~\{ c_i\}_{i \in \mathbb{N}} \textrm{ s.t.} \int_{0}^{\frac{k}{n} t +s} f(v)~dv=\int_{0}^{c_k} g(c)~dc= \int_{0}^{\frac{k}{n} t + s}\int_{0}^{c_k} \hat{h}(v,c) ~dv ~dc , \\ \forall k \in \{ 1, \dots, n\}, \forall (t,s) \in \big\{(0,p^*), (p^*, y(p^*)-p^*), (y(p^*), \overline{v}-y(p^*)) \big\}\bigg\}.
\end{multline*}
 We use $\hat{h}_n \in \mathcal{F}(p^*)$ to denote the joint probability density with $n$ partitions in each of three rectangles.\\
 
\begin{figure}[h]
 \centering
\includegraphics[scale=1.2]{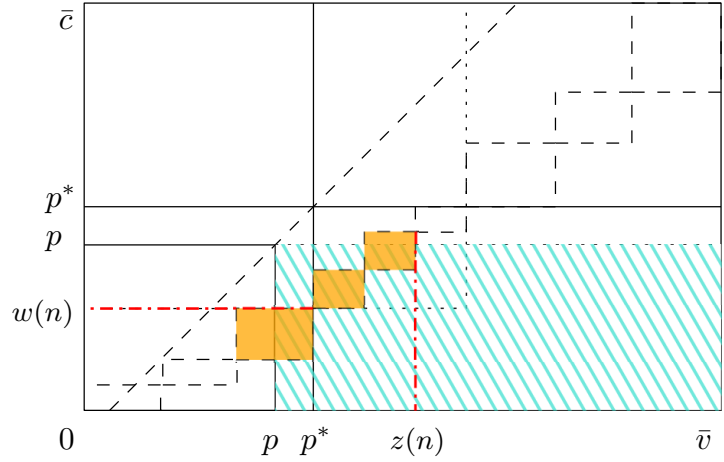}
\caption{Sequence of joint distributions}
\label{fig:seq}
\end{figure}

An element of collection is depicted in Figure \ref{fig:seq}.\\

Notice that there are sets $\{0, v_1, \dots,   v_n\}$ and $\{0, c_1, \dots, c_n  \}$ such that there is mass only in rectangles of form $[v_{k-1}, v_k]\times[c_{k-1}, c_k]$ where $k \in \{ 1, \dots, n\}.$\\

We will show that arbitrary mechanism $M^p$ generates robust efficiency gains less than $A$.

\noindent Consider an arbitrary joint probability density, $\hat{h}_n(v,c) \in \mathcal{F}(p^*). $ From equation~\eqref{eq:gains}, we know that just the marginal density in the region of trade matters for calculation of efficiency gains. Thus, the efficiency gains for posted price mechanism- $ M^p$ is upper bounded by 
\begin{align*}
 \int_p^{z(n)}& vf(v) dv  -  \int_{w(n) }^p cg(c) dc,
 \end{align*}
  where
  \begin{align*}
  w(n)&= \inf \left\{ c_k:  \int_0^p f(v) dv \le \int_0^{c_k} g(c)~dc , ~k \in \{ 1, \dots, n\}\right\}, \\
  z(n)&= \inf \left\{ v_k: \int_0^p g(c) dc \le \int_0^{v_k} f(v) dv ,~ k \in \{ 1, \dots, n\}\right\}.
  \end{align*}
The Figure~\ref{fig:seq} depicts $w(n)$ and $z(n)$ for an example.\\
Note that 
 \begin{align*}
 \int_{0}^{p + \delta_1(n)} f(v)~ dv &= \int_0^{w(n) } g(c)~dc,\textrm{ and}\\
 \int_0^{z(n) } f(v) ~ dv &= \int_0^{p+\delta_2(n)} g(c)~dc.
 \end{align*}
 Since $ \delta_1$ and $\delta_2$ are strictly decreasing in $n$, we have
 \begin{align*}
 w(n)  \rightarrow x(p), &~ z(n)  \rightarrow y(p), ~\textrm{ and} \\
  \int_p^{ z(n) } vf(v) dv -  \int_{w(n) }^p cg(c) dc &\rightarrow  \int_p^ {y(p)} vf(v) dv -  \int_{x(p)}^p cg(c) dc \le A.
  \end{align*}
The efficiency gains of posted price mechanism $M^p$ is given by
  \begin{align*}
  \lim_{n  \rightarrow \infty}   \mathbb{E}_{M^p, \hat{h}_n \in  \mathcal{F}(p^*)} \left(  \int_p^{ z(n) } vf(v) dv -  \int_{w(n) }^p cg(c) dc   \right) \le A.
  \end{align*}
Since the above relation holds for arbitrary posted price mechanism, we get
  \begin{align*}
 \sup_{(q,t)\in \mathcal{M}_P} \inf_{\hat{h} \in \mathcal{F}(p^*)}\int_{v} \int_{c} (v-c) \hat{h}(v,c) ~q(v,c)~dc ~dv \le A.
 \end{align*}
 As $\mathcal{F}(p^*) \subset \mathcal{H}^d$, we get
 \begin{align*}
   \textsc{Eff}(\mathcal{M}_P)&=\sup_{(q,t)\in \mathcal{M}_P} \inf_{\hat{h} \in \mathcal{H}^d}\int_{v} \int_{c} (v-c) \hat{h}(v,c) ~q(v,c)~dc ~dv \le A
  \end{align*}
  However, we have
  \begin{align*}
   \textsc{Eff}(\mathcal{M}_P)&\ge    \textsc{Eff}(M^{p^*})=A.
  \end{align*}
  Thus, we get $   \textsc{Eff}(\mathcal{M}_P) =   A$ and $M^{p^*}$ is an optimal robust mechanism in $\mathcal{M}_P.$
  \linebreak

\subsection{Proof of Theorem \ref{theo:equiv1}}
By definition of $\inf$ and the fact that $\hat{h}_n(v,c)\in \mathcal{F}(p^*)$
\begin{align}
\label{eq:equiv1} \inf_{\hat{h} \in \mathcal{F}(p^*)} \sup_{(q,t)\in \mathcal{M}_P} \int_{v} \int_{c} (v-c) \hat{h}(v,c) ~q(v,c)~dc ~dv&\le \lim_{n \rightarrow \infty} \sup_{(q,t)\in \mathcal{M}_P} \int_{v} \int_{c} (v-c) \hat{h}_n(v,c) ~q(v,c)~dc ~dv.
\end{align}
By optimality of $M^{p^*}$ for very fine joint distributions proved in Theorem \ref{theo:rob},
\begin{align}
\nonumber \lim_{n \rightarrow \infty} \sup_{(q,t)\in \mathcal{M}_P} \int_{v} \int_{c} (v-c) \hat{h}_n(v,c) ~q(v,c)~dc ~dv&=   \lim_{n  \rightarrow \infty} \int_{v>p^*} \int_{c<p^*} (v-c)~\hat{h}_n(v, c)~ dc ~dv. \\
\label{eq:equiv2}&= A
\end{align}
% The last equality follows from the fact that for any joint probability density, $\hat{h}_n(v,c) \in \mathcal{F}(p^*)$, the efficiency gains is $A$ as it is of the form of worst distribution for mechanism $M^{p^*}.$
 Combining (\ref{eq:equiv1}) and (\ref{eq:equiv2}) and the fact $\mathcal{H}^d \supset \mathcal{F}(p^*)$ that we get
\begin{align}
\label{eq:less1} \inf_{\hat{h} \in \mathcal{H}^d} \sup_{(q,t)\in \mathcal{M}_P} \int_{v} \int_{c} (v-c) \hat{h}(v,c) ~q(v,c)~dc ~dv&\le A.
 \end{align}
However,
 \begin{align}
\nonumber \inf_{\hat{h} \in \mathcal{H}^d} \sup_{(q,t)\in \mathcal{M}_P} \int_{v} \int_{c} (v-c) \hat{h}(v,c) ~q(v,c)~dc ~dv& \ge \sup_{(q,t)\in \mathcal{M}_P}\inf_{\hat{h} \in \mathcal{H}^d} \int_{v} \int_{c} (v-c) \hat{h}(v,c) ~q(v,c)~dc ~dv\\
\label{eq:more1} &\equiv A.
\end{align}
Equations (\ref{eq:less1}) and (\ref{eq:more1}) establishes equivalence between min-max and max-min exercise.

\end{document}